\documentclass{aastex}          
\usepackage{spr-astr-addons}    

\usepackage{url}
\usepackage{soul}
\soulregister{\cite}7 
\soulregister{\citep}7 
\soulregister{\citet}7 

\RequirePackage{color}
\begin{document}
%
\title{Investigation and Application of Fitting Models for Centering Algorithms in Astrometry}

\shorttitle{Fitting Models for the Centering Algorithms in Astrometry}
\shortauthors{Lin et al.}

\author{F. R. Lin\altaffilmark{1,2,3}}
\and
\author{Q. Y. Peng\altaffilmark{1,3}}
\and
\author{Z. J. Zheng\altaffilmark{1,3}}
\and
\author{B. F. Guo\altaffilmark{1,3}}
\and
\author{Y. J. Shang\altaffilmark{1,3}}
\email{tpengqy@jnu.edu.cn} 

\altaffiltext{1}{Sino-French Joint Laboratory for Astrometry, Dynamics and Space Science, Jinan University,
Guangzhou 510632, China. {\it tpengqy@jnu.edu.cn}}
\altaffiltext{2}{School of Software, Jiangxi Normal University, Nanchang 330022, China.}
\altaffiltext{3}{Department of Computer Science, Jinan University, Guangzhou 510632, China.}

\begin{abstract}
To determine the precise positions of stars in CCD frames, various centering algorithms have been proposed for astrometry. The effective point spread function (ePSF) and the Gaussian centering algorithms are two representative centering algorithms. This paper compares in detail and investigates these two centering algorithms in performing data reduction. Specifically, synthetic star images in different conditions (i.e. profiles, fluxes, backgrounds and full width at half maximums) are generated and processed. We find that the difference in precision between the two algorithms is related to the profiles of the star images. Therefore, the precision comparison results using an ideal Gaussian-profile star image cannot be extended to other more specific experimental scenarios. Based on the simulation results, the most appropriate algorithm can be selected according to the image characteristics of observations, and the loss of precision of other algorithms can be estimated. The conclusions are verified using observations captured by the 1-m and 2.4-m telescopes at Yunnan Observatory.
\end{abstract}


\section{Introduction}           
\label{sect:intro}

Determining the center of a star image precisely is fundamental to high-precision astrometry. With the release of Gaia DR2 \citep{brown2018gaia} and the development of astrometric techniques \citep{morgado2019approx,wang2019distortion,lin2019characterization,lin2020Using}, the influences of other error components in astrometry, such as geometric distortion and atmospheric turbulence, are decreasing. As a result, the importance of centering precision becomes more and more prominent.

To compare the precision of various centering algorithms, many comparisons have been performed by researchers \citep{van1975digital,chiu1977astrometric,auer1978digital,stone1989comparison,platais1991astrometry,lu1993digital}. Among them, two-dimensional Gaussian fitting is considered to have the highest precision in most conditions. However, when a star image is undersampled (FWHM$<$2 pixels), high precision position cannot be obtained using traditional centering algorithms \citep{stone1989comparison}. This arises due to incorrect point spread function (PSF) fitting which introduces a so-called pixel-phase error. The effect is even more serious in Hubble Space Telescope's imaging cameras. \citet[hereafter, called Paper 1]{anderson2000toward} proposed the effective PSF (ePSF) algorithm to address this problem. This method works well for undersampled images, because the source of the pixel-phase error can be eliminated when an accurate ePSF is constructed. Using the ePSF algorithm, observations taken with Hubble Space Telescope can be measured with high accuracy of 0.005 pixels \citep{piatek2002proper}. The ePSF algorithm is also adapted to well-sampled ground-based astrometry \citep{anderson2006psfs}. On the other hand, the pixel-phase error can be corrected by some techniques independent of the fitting process as well \citep{zacharias2017ucac5}, and Gaussian fitting does not introduce the pixel-phase error for well-sampled images. Coupled with some historical reasons \citep{espinosa2018optimality}, Gaussian fitting has become the method commonly used in ground-based CCD astrometry to obtain high-precision results \citep[e.g.][to name a few]{monet1983ccd,stone2003upgrades,peng2017precise,zacharias2017ucac5}. Namely, the two algorithms above can be selected depending on the preference when performing data reduction, and the difference in precision between them is to be investigated.


When star images have an ideal Gaussian profile, \citet{lobos2015performance} characterized the difference between the least squares estimator (i.e. Gaussian fitting in this context) and the minimum variance Cram$\rm \acute{e}$r-Rao (CR) bound analytically. Moreover, the performance of the adaptive weighted least squares estimator is shown to be very close to the CR bound in all observational regimes \citep{espinosa2018optimality}. The ePSF fitting adopts the same weight as the latter estimator and therefore could have the same precision. In other words, the theoretical prediction can well-estimate the difference between the Gaussian and the ePSF fitting in the ideal situation. The variance of the Gaussian fitting is as large as $8/(3\sqrt3)$ times of the ePSF fitting for a well-sampled star image with an ideal high signal-to-noise ratio (SNR), while their variances are almost the same for low SNR star images \citep{lobos2015performance}. Our simulations also show agreements with the prediction.

However, the profile of practical observations is usually deviated from the ideal Gaussian profile, which makes the conclusion of the precision comparison in practical situations different from the theoretical prediction.
In order to investigate the precision of each algorithm in practical application, a series of synthetic star images under different observational conditions are generated for the simulations. The conditional variables include the profile of stars, flux, background noise and the FWHMs of star images.
The spatial variation of the PSF across the detector is also taken into account.
We not only use the Gaussian profile as \citet{stone1989comparison} did, but also use the Moffat profile \citep{moffat1969a} and the practical ePSF extracted from observations to generate synthetic images. With these profiles introduced, we can draw practical conclusions from the simulations. The synthetic images are fitted with a combination of different fitting models and weighting schemes to get the simulation results. Finally, the conclusions of this work are verified by two sets of observations obtained from the 1-m and 2.4-m telescopes at Yunnan Observatory. Using a statistical method mentioned in \citet{lin2019characterization}, the centering precision of practical observations can be shown clearly.

The contents of this paper are arranged as follows. The second section briefly introduces the processes of the Gaussian and the ePSF algorithms. A detailed description of the synthetic images and observations is shown in Section \ref{expImg}. In the fourth section, the centering results of the algorithms are compared and analyzed. In addition, the application of these algorithms in the reduction of the practical observations are given. The last section provides conclusions.
\section{Methods}\label{meds}
\subsection{Basic principle}\label{sec2}
The ePSF, which is proposed by Paper 1, is a convolution that represents the total fraction of a star's light falling in the area of each pixel, and it describes the profile of a star image directly.

Let the ePSF of a star image be $p$, and $p(x,y;x_{c},y_{c})$ expresses the value of the function $p$ at pixel $(x,y)$ when the center of $p$ is $(x_{c},y_{c})$.
According to the method of maximum likelihood \citep{cowan1998statistical}, the center $(x_{c},y_{c})$ and the total flux $f_{*}$ of a star image can be solved by minimize the quantity
\begin{small}
\begin{equation}
\label{eq6p}
\chi^2(x_{c},y_{c},f_{*})=\sum_{(x,y)\in S}{\frac{\left[M_{x,y}- f_{*}\cdot p(x,y;x_{c},y_{c})-B\right]^2}{M_{x,y}}},
\end{equation}
\end{small}
where $B$ is the background of the star image, $M_{x,y}$ the actual count level in pixel $(x,y)$, and $S$ the set of pixels used in the centering process. Equation (\ref{eq6p}) here corresponds to Equation (10) in Paper 1. If the noise of the star image is uniform everywhere, the equation can be reduced to
\begin{small}
\begin{equation}
\label{eq8}
\chi^2(x_{c},y_{c},f_{*})=\sum_{(x,y)\in S}{\left[M_{x,y}- f_{*}\cdot p(x,y;x_{c},y_{c})-B\right]^2}.
\end{equation}
\end{small}

Obviously, the most important and challenging task in the centering process is to find an accurate function $p$. The Gaussian centering algorithm simply uses a two-dimensional symmetric Gaussian function as $p$, but the ePSF algorithm can construct an accurate ePSF to fit star images. Since the PSF of observations is usually not an ideal Gaussian function, the precision of the ePSF fitting can be higher than that of the Gaussian fitting to varying degree according to observational conditions. A detailed comparison will be shown by the simulations given in Section~\ref{Reses}.

\subsection{The Gaussian centering algorithm}\label{GaussFit}
Although several variants of the Gaussian fitting have been developed, the best astrometric results are usually obtained from fitting a symmetric Gaussian function plus a flat background to the image \citep{auer1978digital}. Therefore, the function $p$ used in the Gaussian fitting is expressed as:
\begin{small}
\begin{eqnarray}
\label{eq9}
p_{{}_{G}}(x,y;x_{c},y_{c})=\alpha\ {\rm exp}\left[-\frac{(x-x_{c})^2+(y-y_{c})^2}{2\sigma^2} \right] +\beta,
\end{eqnarray}
\end{small}
where $\alpha$ is the amplitude, $\beta$ is the baseline, $(x_{c},y_{c})$ and $\sigma$ are the center and rms half-width of the star image, respectively. In fact, the parameter $\beta$ would be 0 provided that the background of the star image $B$ is determined accurately.

In general, Gaussian fitting uses Equation~(\ref{eq8}) in the least squares fitting process. When Equation~(\ref{eq6p}) is adopted, the solution is referred to as $weighted\ Gaussian$ fitting in this paper. Note that the weight used in the Gaussian or the weighted Gaussian fitting (i.e. unit weight or 1/$M_{x,y}$) is not the optimal weight, and a description of the optimal weight was given in \citet{espinosa2018optimality}.

\subsection{The effective PSF algorithm}

The process of the ePSF algorithm will only be briefly introduced here. Detailed description has been given in Paper 1 and some other papers \citep[e.g.][]{piatek2002proper,anderson2006psfs,anderson2006ground}.
The construction of the ePSF requires at least dozens of sampling stars whose centers are located in different parts of the pixel, with different pixel-phase. The ePSF value at each position is calculated by the basic equation
\begin{eqnarray}
\label{eq10}
p_{{}_{E}}(x,y;x_{c},y_{c})=\frac{M_{x-x_c,y-y_c}-B}{f_{*}}.
\end{eqnarray}
 Parameters used in the construction of the ePSF, such as the sampling range of the ePSF and the size of the smoothing kernel, are calculated automatically based on the mean FWHM of the star images in a CCD frame. The smoothing kernels with different sizes are generated using the algorithm proposed by \citet{kuo1991multidimensional}. After the ePSF of the frame is constructed, the center of each star in the frame is solved by Equation~(\ref{eq6p}).

In order to study the effect of the model construction error on the centering precision, the true ePSF, which is used to generate synthetic images, will also be used as a fitting model in our simulations. In this case, the method is referred to as the $true\ ePSF$ fitting hereafter.

In our program, the radius $R_{f}$ of the circular area used in the centering process is 2.5$\sigma$ by default, because other factors that cut down the centering precision (such as the flux of adjacent stars and cosmic ray in practical observations) may be introduced when $R_{f}$ is too large.
\section{Experimental images}\label{expImg}

\subsection{Synthetic Images}\label{synImg}
Since the centering precision is mainly affected by the FWHM, background and flux of a star image \citep{king1983accuracy,mendez2013analysis,mendez2014analysis}, only these three factors have been considered in previous works. In this paper, another factor, namely the effect of profiles on the difference in precision between algorithms, is also investigated.

There are five profiles used in the simulations, corresponding to five sets of synthetic images.

1. The profile of the two-dimensional symmetric Gaussian function which is expressed by
\begin{equation}
\label{eq11}
F(x,y)={\rm exp}\left[-\frac{(x-x_{c})^2+(y-y_{c})^2}{2\sigma^2} \right].
\end{equation}

2. The Moffat profile with very sharp center and gentle wings. The profile corresponds to the two-dimensional Moffat function (Moffat 1969)
\begin{equation}
\label{eq12}
F(x,y)=\left[ 1+\frac{(x-x_c)^2+(y-y_c)^2}{r_0^2}\right]^{-\beta},
\end{equation}
where $r_0$ is a parameter controlling the width of the profile and $\beta$ affects the shape of the profile. Since this profile is used to investigate the effect of the profile in the extreme case, an atypical value of 1.2 for $\beta$ is used in our simulations.

\begin{figure}

	\centering
\includegraphics[width=0.32\textwidth]{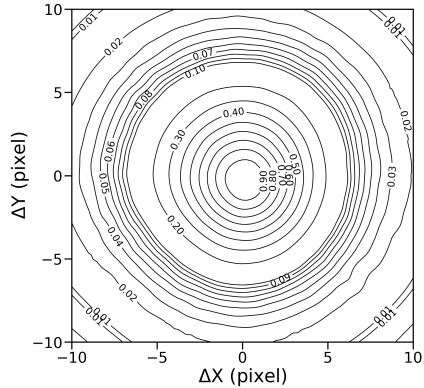}
		\caption{Contour lines of an ePSF which is constructed from one frame of the observations taken with the 1-m telescope at Yunnan Observatory. In the construction of the ePSF,
each pixel is divided into 4 grid points. The ePSF has been normalized so that its peak is 1.} 	
	\label{fig1}
\end{figure}

3. The single ePSF which is constructed from one frame of the observations taken with the 1-m telescope at Yunnan Observatory. Figure~\ref{fig1} shows the contour lines of the ePSF.

\begin{figure}	
	\centering
\includegraphics[width=0.32\textwidth]{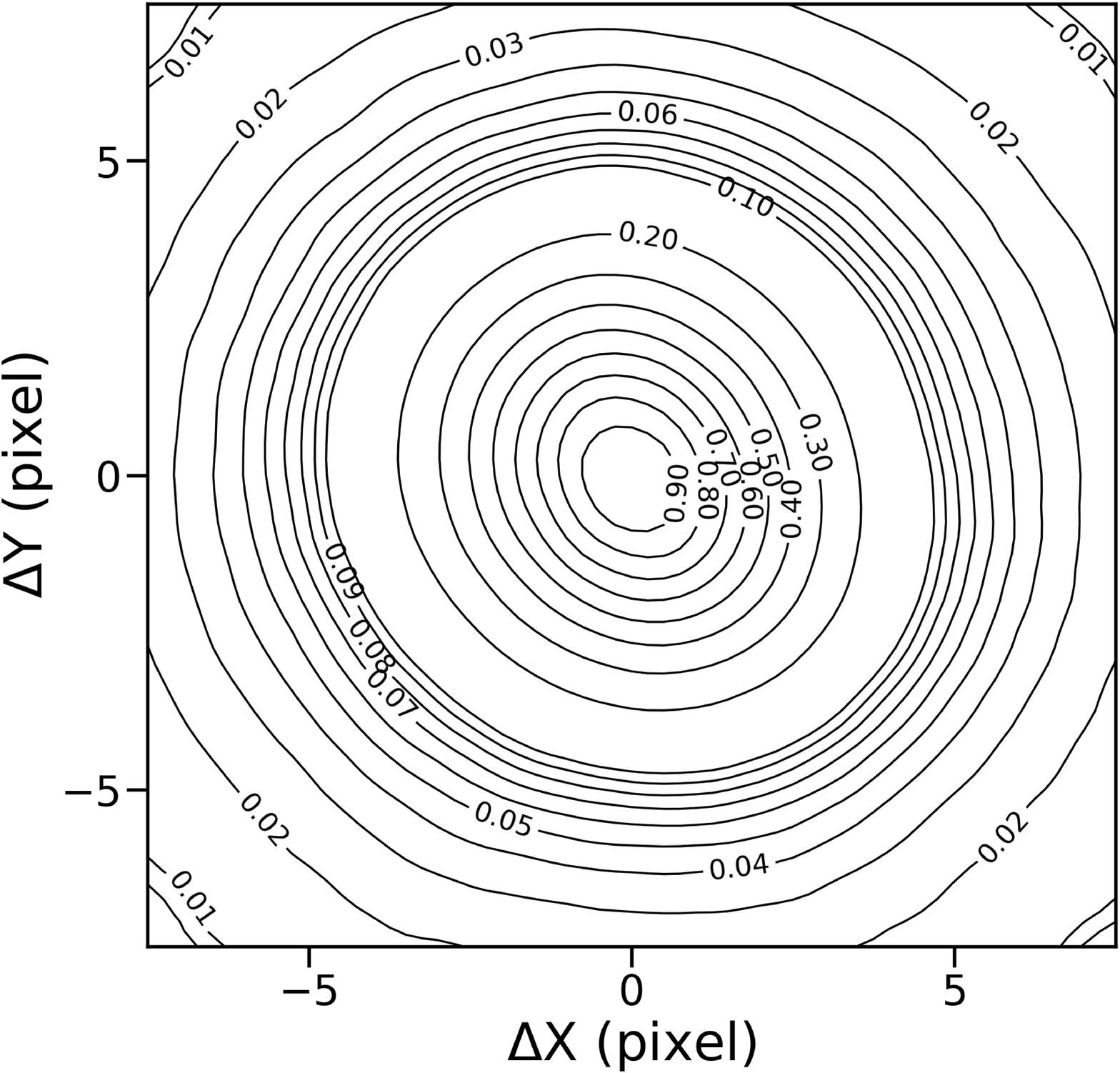}
		\caption{Contour lines of an asymmetric ePSF which is constructed from a poor focusing frame of the observations taken with the 2.4-m telescope at Yunnan Observatory.} 	
	\label{fig4}
\end{figure}

4. An asymmetric ePSF which is constructed from a poor focusing CCD frame captured by the 2.4-m telescope at Yunnan Observatory. The contour lines of this ePSF are shown in Figure \ref{fig4}.

\begin{figure}	
	\centering
\includegraphics[width=0.35\textwidth]{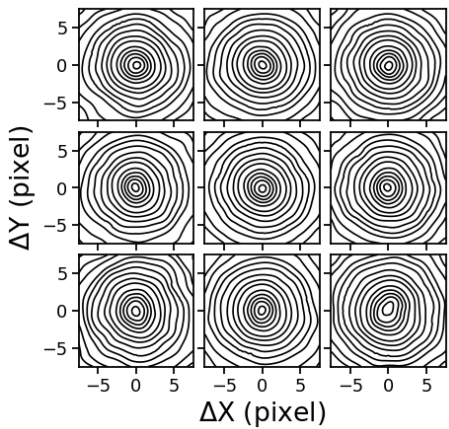}
		\caption{Contour lines of a $3\times3$ ePSFs that constructed from one frame of the observations taken with the 2.4-m telescope at Yunnan Observatory. The contour levels are adjusted to show the ePSFs more clearly.} 	
	\label{fig2}
\end{figure}

5. The multiple ePSFs which are constructed from one frame of the 2.4-m telescope observations. The ePSF at any given position can then be interpolated using these fiducial ePSFs. Figure \ref{fig2} shows the contour lines of these fiducial ePSFs.

\begin{figure*}	
	\centering
\includegraphics[width=0.32\textwidth]{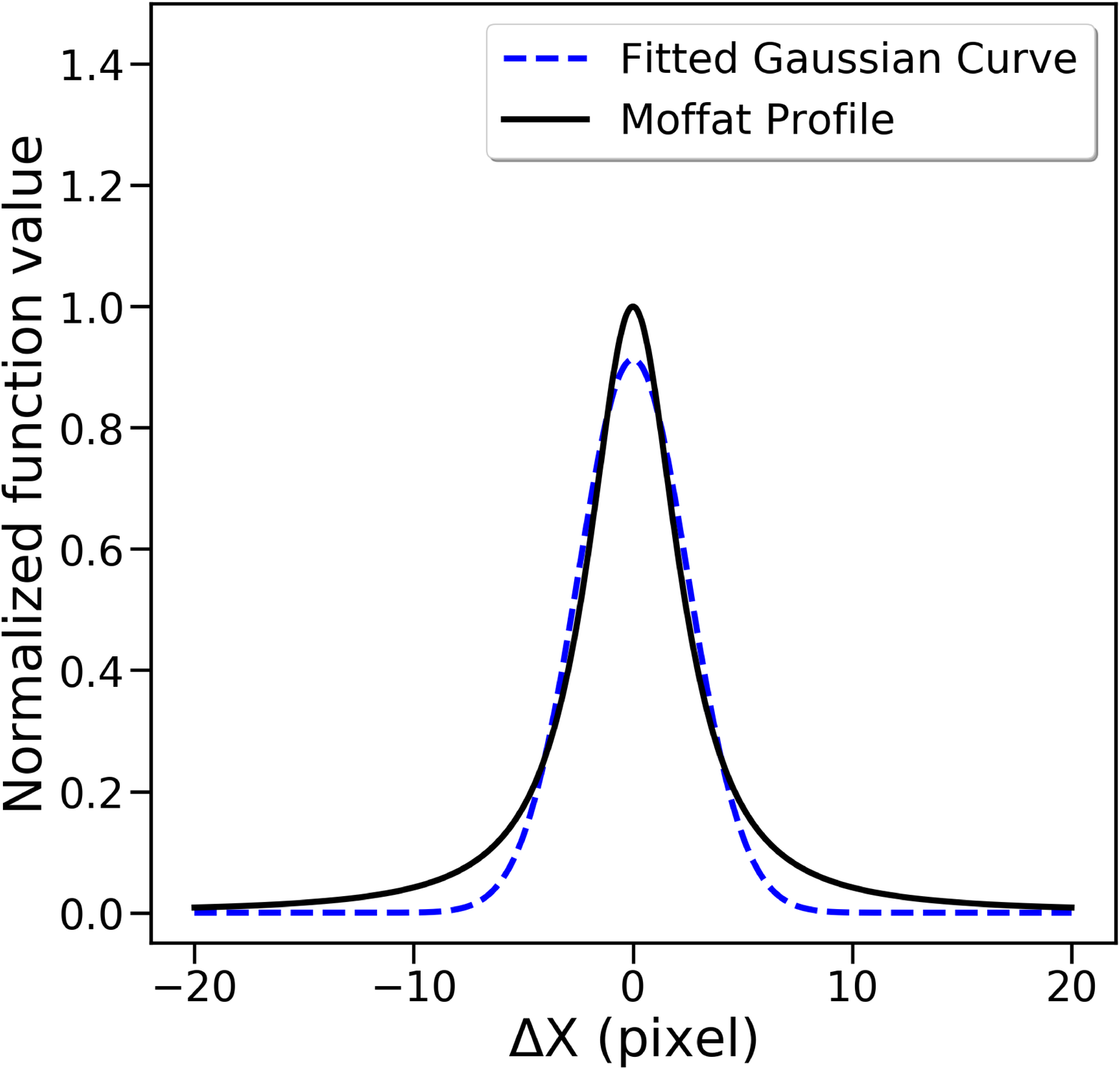}
\includegraphics[width=0.32\textwidth]{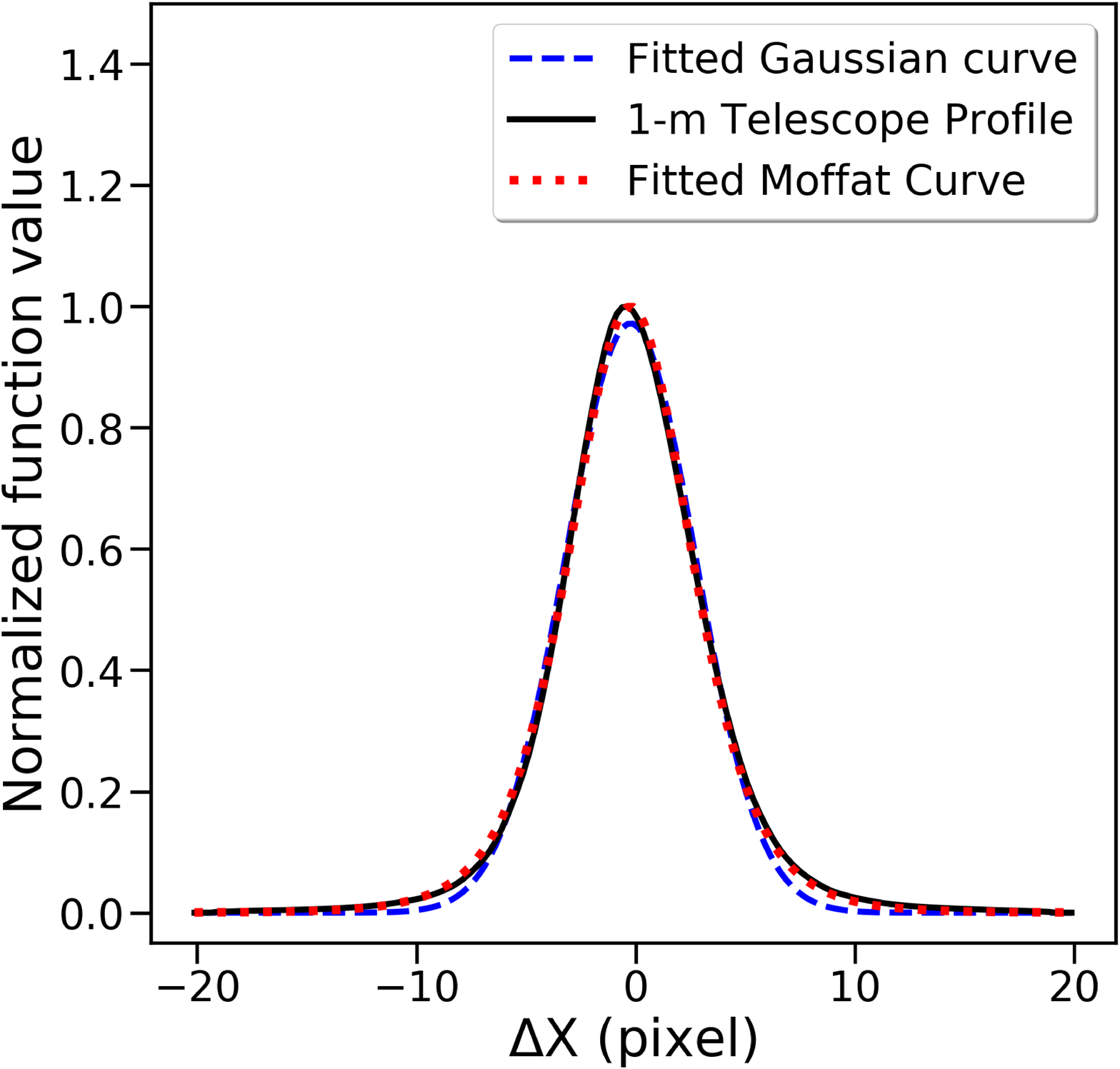}
\includegraphics[width=0.32\textwidth]{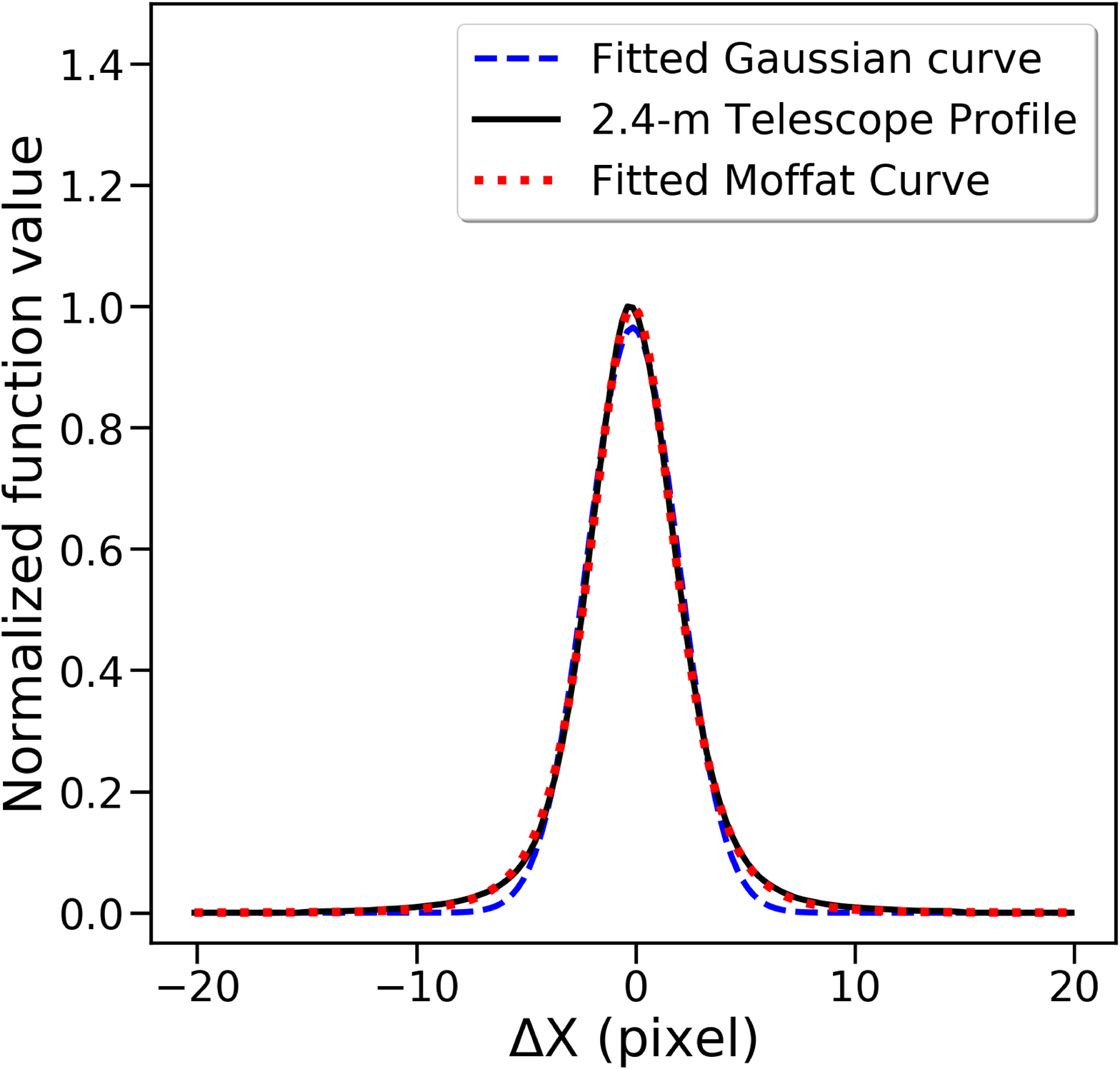}
		\caption{The solid line is the $x$ profile through the center of the ePSF along the $x$ direction, the dashed line the corresponding Gaussian curve that fits it exactly. The dotted lines in the middle and right panels are the Moffat curves that fit the practical profiles exactly. Left panel shows the Moffat profile when parameters $r_0=11.0$ and $\beta=1.2$. Middle and Right panels show Profiles 3 and 4, respectively.} 	

	\label{fig3}
	
\end{figure*}

The Moffat profile is chosen since we find that the practical ePSF usually have a sharper center than the Gaussian function, and so fitting it with the Moffat function is better than the Gaussian. Furthermore, the practical ePSF includes more factors that can affect the centering precision, such as the asymmetry of profiles.
Figure \ref{fig3} shows Profiles 2, 3, 4 and corresponding Gaussian curves that fit them exactly, the practical Profiles 3 and 4 are also fitted with Moffat curves. As shown in the figure, it is obvious that the practical profiles are more consistent with Moffat curves than with Gaussian. The fitted Moffat parameter $\beta$ is 3.5 for Profile 3 and 2.8 for Profile 4.
Note that when the parameter $\beta$ of Moffat function approaches infinity, the Moffat profile becomes Gaussian.
Therefore, the analytical Profiles 1 and 2 can be considered as two extreme cases of practical profile changes.

\begin{figure}	
	\centering
\includegraphics[width=0.26\textwidth]{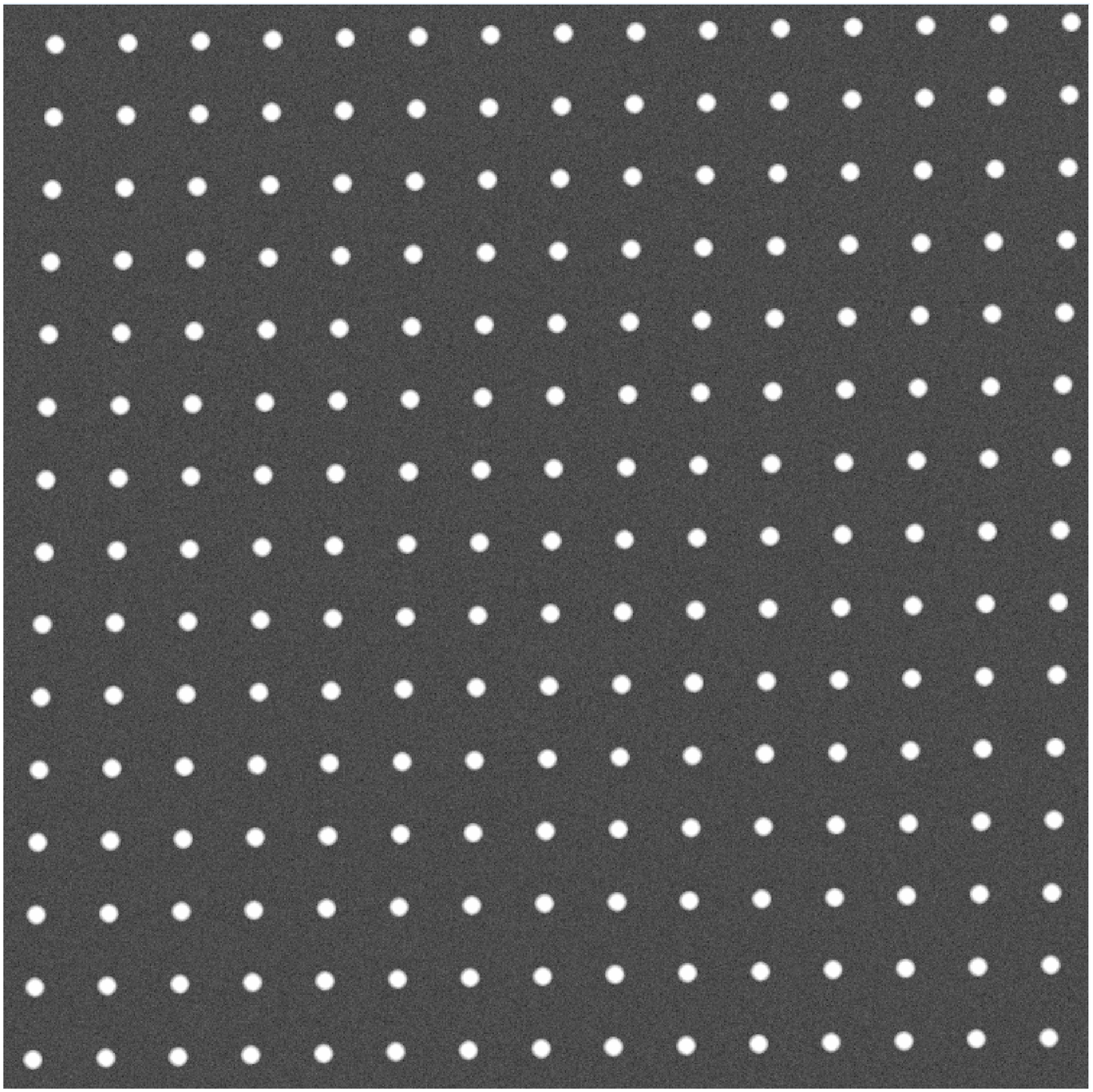}
		\caption{One synthetic frame in simulation set 3. The total flux of each star in a frame is the same and the pixel phase of each star is random. The star images are arranged as shown in the figure to ensure that each star is absolutely isolated.} 	
	\label{fig5}
\end{figure}

The noise model proposed by \citet{merline1995realistic}, which contains the readout noise of Gaussian distribution and other noises of Poisson distribution, is used as a criterion for adding noises to the image.
Each of the synthetic frames contains isolated stars with same brightness (i.e. total flux), and the brightness of the stars varies with different frames. The peak values of the star images range from just detectable to about 60000 counts. Table \ref{tbl1} shows the detailed information of the five sets of simulations and Figure \ref{fig5} shows one synthetic frame in simulation set 3.

\begin{table*}
\begin{center}
\caption{Detailed information of the five sets of synthetic frames used in this paper. Column 2 denotes the profile used to generate the images. Column 3 is the standard deviation (std) of the 2-D Gaussian model which fits the star images exactly, and it represents the width of the profile. Column 4 is the approximate std of the background noise. Column 5 gives the number of stars in each synthetic frame. All the frames have the same width and height of 2048 pixels.\label{tbl1}}
\begin{tabular}{@{}l*{15}{l}}
\tableline
ID&Profile&$\sigma_{_{P}}$&$\sigma_{_{B}}$&Stars\\
&&(pixel)&(counts)&\\
\tableline
1&Gaussian &3.0                &1,10,100&225 \\
2&Moffat   &1.2, 1.5, 3.0     &1,10,100&225   \\
3&1-m telescope&3.0                &1,10,100&225   \\
4&asymmetric   &2.1                &1,10,100&225   \\
5&2.4-m telescope&$2.0$    &1, 10    &360   \\
\tableline
\end{tabular}
\end{center}
\tablecomments{Only the approximate std of the background in each pixel is listed in the table, and the excepted value is simply set to 1000 to make the count level in each pixel positive. Since the full width at half maximums of the 3$\times$3 ePSFs are not very different, only the average $\sigma_{_{P}}$ of them is listed.}
\end{table*}

\subsection{Observations}\label{obsImg}
Two sets of observations are used to verify the conclusions of simulations.
Observation set 1 contains 54 frames of CCD images captured by the 1-m telescope at Yunnan Observatory (IAU code 286, longitude\---E$102^{\circ}47^{\prime}18^{\prime\prime}$, latitude--N$25^{\circ}1^{\prime}30^{\prime\prime}$, and height--2000m above sea level) on February 24, 2011. We use observations obtained many years ago because the environment of the observatory changed after that time, which led to the increase of the sky background. Besides, a new CCD had just been installed in 2009, so it had better performance at that time. Observation set 2 contains 44 frames of CCD images captured by the 2.4-m telescope (IAU code O44, longitude\---E$100^{\circ}1^{\prime}51^{\prime\prime}$, latitude--N$26^{\circ}42^{\prime}32^{\prime\prime}$, and height--3193m above sea level) at Yunnan Observatory on January 3, 2011. The targets of these two sets of observations are open cluster M35 and NGC1664 respectively. Detailed information about these observations is given in Table \ref{tbl2}. More instrumental details of the reflectors and CCD detectors are listed in Table \ref{tbl3}.

\begin{table*}
\caption{Detail information of the observations captured by the 1-m and 2.4-m telescopes at Yunnan Observatory. The value in parentheses is the average. \label{tbl2}}
\begin{center}
\begin{tabular}{@{}l*{15}{l}}
\tableline\tableline
 &1-m telescope&2.4-m telescope\\
\tableline
Observation date (y-m-d)&2011-2-24& 2011-1-3\\
Frames&54 &44 \\
Exposure time & 18-30 s &30-60 s \\
FWHM  & 5.6-8.2 (6.9) pixels & 4.3-5.9 (5.1) pixels\\
Saturate counts&about 63000 counts& about 700000 counts\\
Background std  & 5.1-6.8 (6.4) counts& 47.2-61.9 (51.5) counts\\
Stars per frame & 259-373 (323)& 824-1148 (945)\\
Fitted Moffat parameter $\beta$& 2.9-4.8 (3.4)& 2.0-2.5 (2.2)\\
\tableline
\end{tabular}
\end{center}
\tablecomments{The parameter $\beta$ is obtained by fitting the constructed ePSF with the 2-D Moffat function. A large $\beta$ indicates that the PSF is close to the Gaussian function.}
\end{table*}

\begin{table*}
\caption{Specifications of the 1-m and 2.4-m telescopes and the corresponding CCD detectors.\label{tbl3}}
\begin{center}
\begin{tabular}{@{}l*{15}{l}}
\tableline\tableline
Parameter&1-m telescope&2.4-m telescope\\
\tableline
Approximate focal length&1330 cm &1920 cm \\
F-Ratio & 13 &8 \\
Diameter of primary mirror&100 cm & 240 cm \\
Approximate scale factor & 0.209 arcsec pixel$^{-1}$ & 0.286 arcsec pixel$^{-1}$\\
Size of CCD array (effective)& 2048 $\times$ 2048&1900 $\times$ 1900 \\
Size of pixel &13.5 $\mu m\ \times$ 13.5 $\mu m$ & 13.5 $ \mu m\ \times$ 13.5 $\mu m$\\
\tableline
\end{tabular}
\end{center}
\end{table*}

\section{Results}\label{Reses}

\subsection{Simulation results}\label{simRes}
\begin{figure*}	
	\centering
\includegraphics[width=0.42\textwidth]{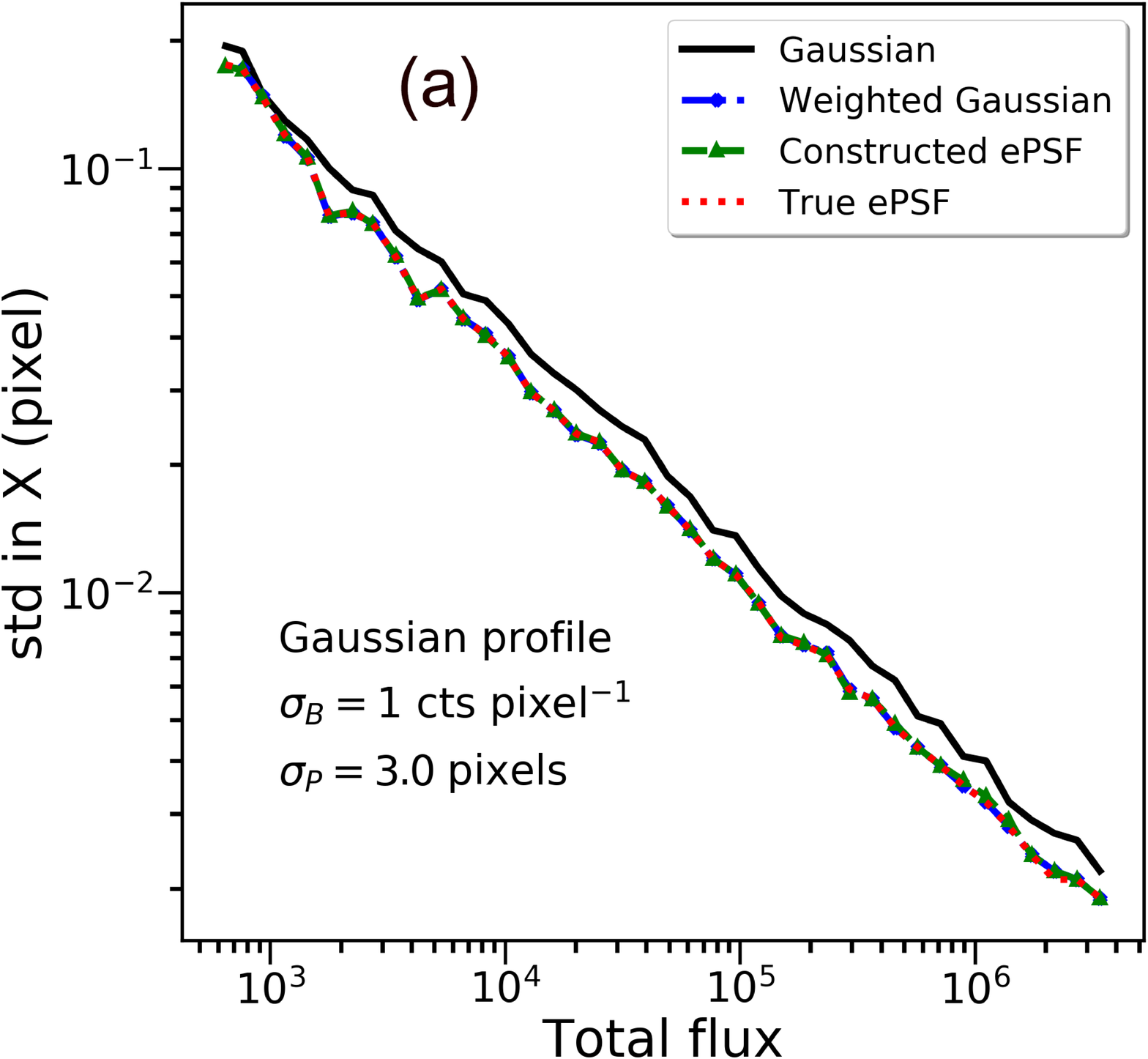} \
\includegraphics[width=0.42\textwidth]{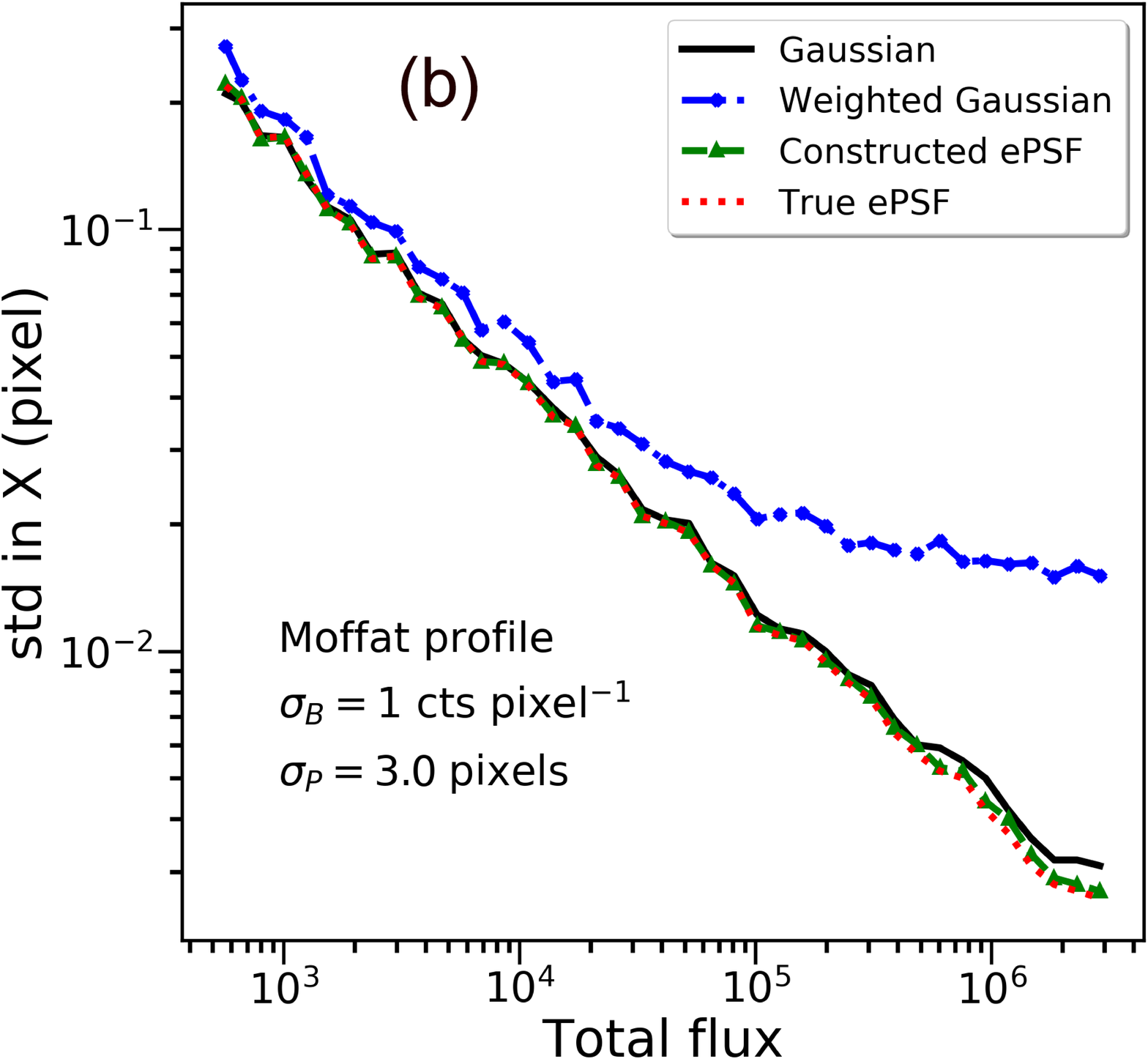}
\includegraphics[width=0.42\textwidth]{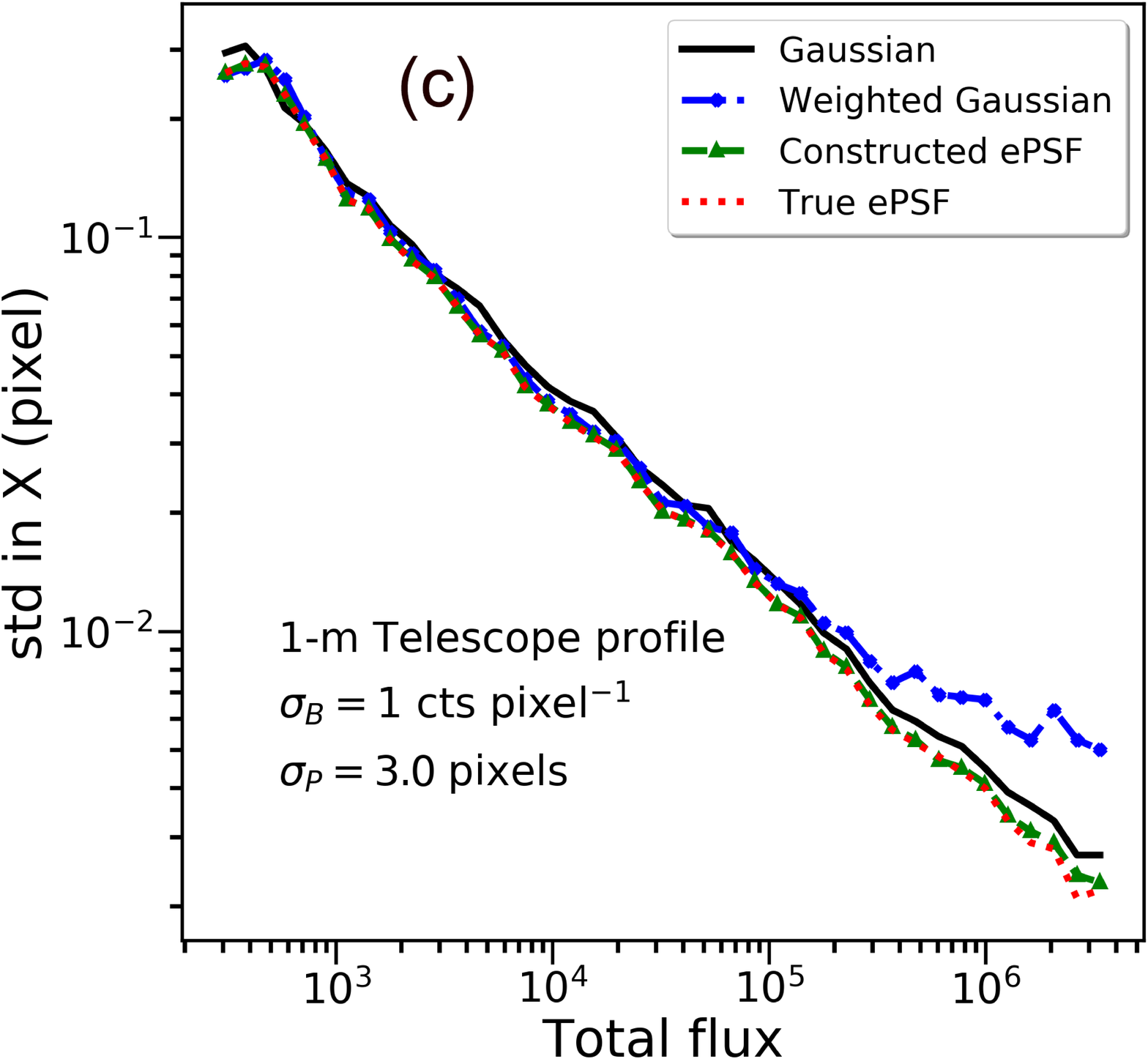}\
\includegraphics[width=0.42\textwidth]{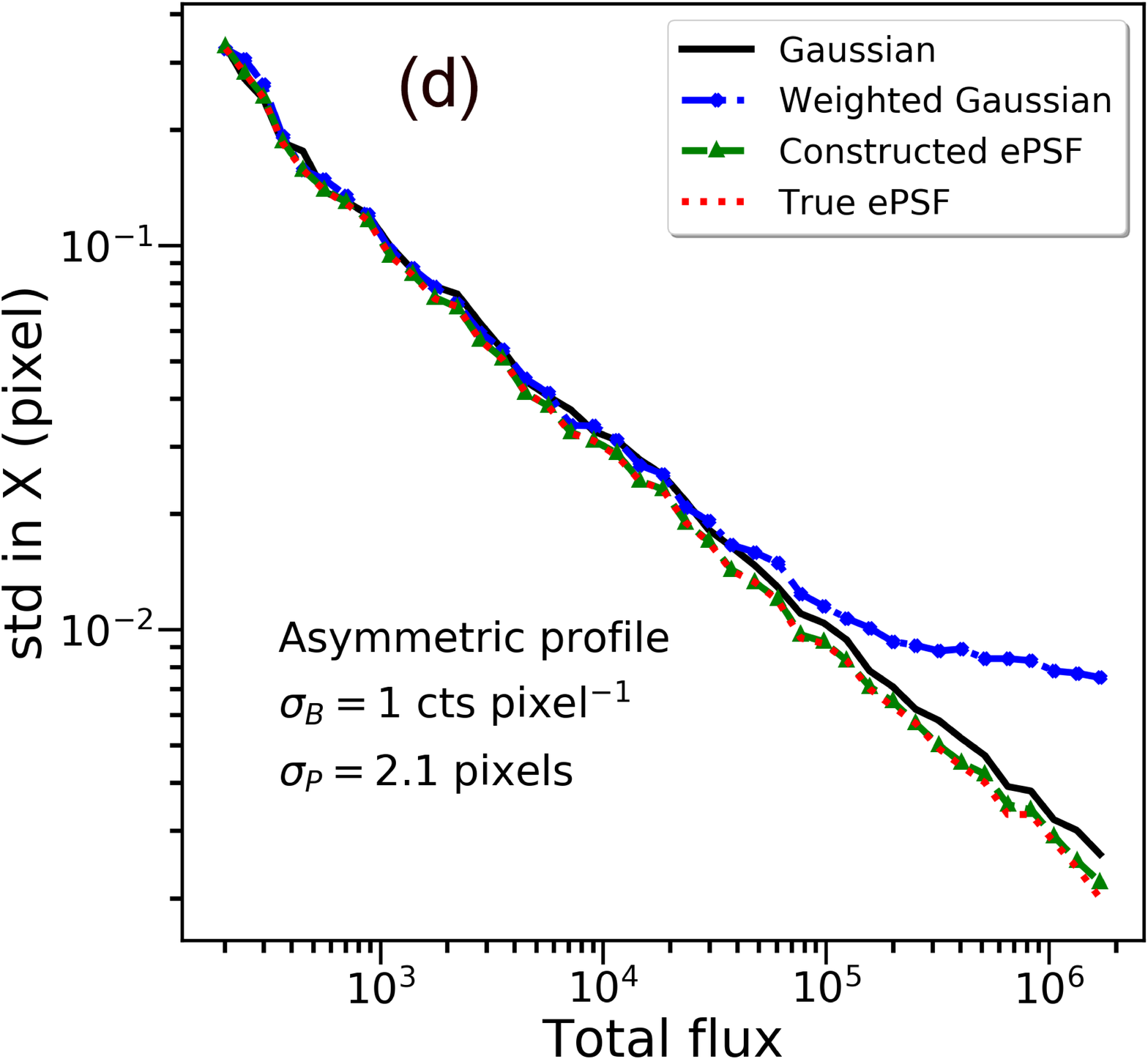}
		\caption{Centering errors of the synthetic star images with different profiles when the background is very low. The vertical axis is the standard deviation (std) of residuals between real and measured positions. Each std is a statistical result of about 225 residuals. The horizontal axis is the total flux of each star. $\sigma_{_{P}}$ is the std of the 2-D Gaussian model which fits the ePSF of the star images exactly, and it represents the width of the profile. $\sigma_{_{B}}$ is the approximate std of the background noise.}
	\label{fig6}
\end{figure*}

Each synthetic star image is centered by the Gaussian fitting, weighted Gaussian fitting, ePSF fitting and the true ePSF fitting, algorithms respectively. A description of them has been given in Section \ref{meds}. For well-sampled star images, each ePSF is sampled from about only 60 stars with random brightness (the brightness distribution is obtained from our observations) to ensure that its model accuracy can be achieved for almost all open cluster observations. In addition, the results of the true ePSF fitting are included to show the maximum precision of the ePSF algorithm, especially for the bright stars. In practical applications, the accuracy of the constructed ePSF is usually not less than that in our simulations, so the centering precision of the true ePSF would be more relevant. The centering precision of the synthetic images with different characteristics is shown in Figures ~\ref{fig6},~\ref{fig8},~\ref{fig11} and ~\ref{fig10}. We only consider the results in the $x$ direction since no coordinate related systematic error exists in the simulations.

\begin{figure}	
	\centering
\includegraphics[width=0.35\textwidth]{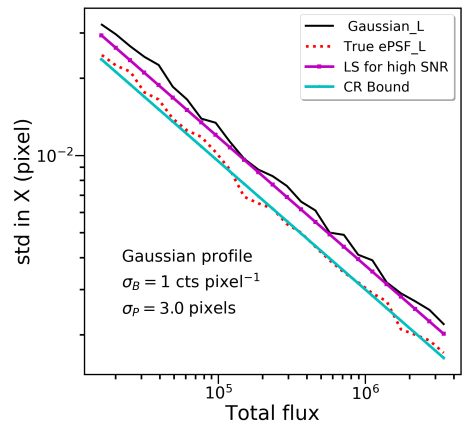}
		\caption{A comparison between the results of different centering algorithms and corresponding theoretical predictions for bright star images with very low background. The suffix $\_$L of the notations in the legend indicates that the method uses a large fitting radius ($R_{f}$=3.5$\sigma$) in the centering process.}
	\label{fig6p}
\end{figure}

The Gaussian function is the most commonly used PSF in the study of centering algorithms. For well-sampled star images with a Gaussian profile, our simulations show that the precision of the ePSF fitting and the weighted Gaussian fitting are obviously higher than that of the Gaussian fitting in high SNR regime (see Figure~\ref{fig6}(a)), and their precision tends to be the same when the SNR is low (the left panel of Figure~\ref{fig8}).
 Figure~\ref{fig6p} shows the square root of CR Bound calculated by the equation \citep{mendez2014analysis}
\begin{equation}
\label{eqcrbound}
\sigma_{x_{c_{1D}}}^{2}\approx\frac{1}{8ln2}\frac{1}{f_*}\rm{FWHM^2}
\end{equation}
for star images whose flux is much higher than the background. The predicted precision of the LS estimator for ideal high SNR regime \citep{lobos2015performance} is also shown in that figure. Compared with the results of the Gaussian fitting and ePSF fitting with enough large fitting radius ($R_{f}$=3.5$\sigma$), our results show good agreement with the theoretical predictions.

\begin{figure*}	
	\centering
\includegraphics[width=0.42\textwidth]{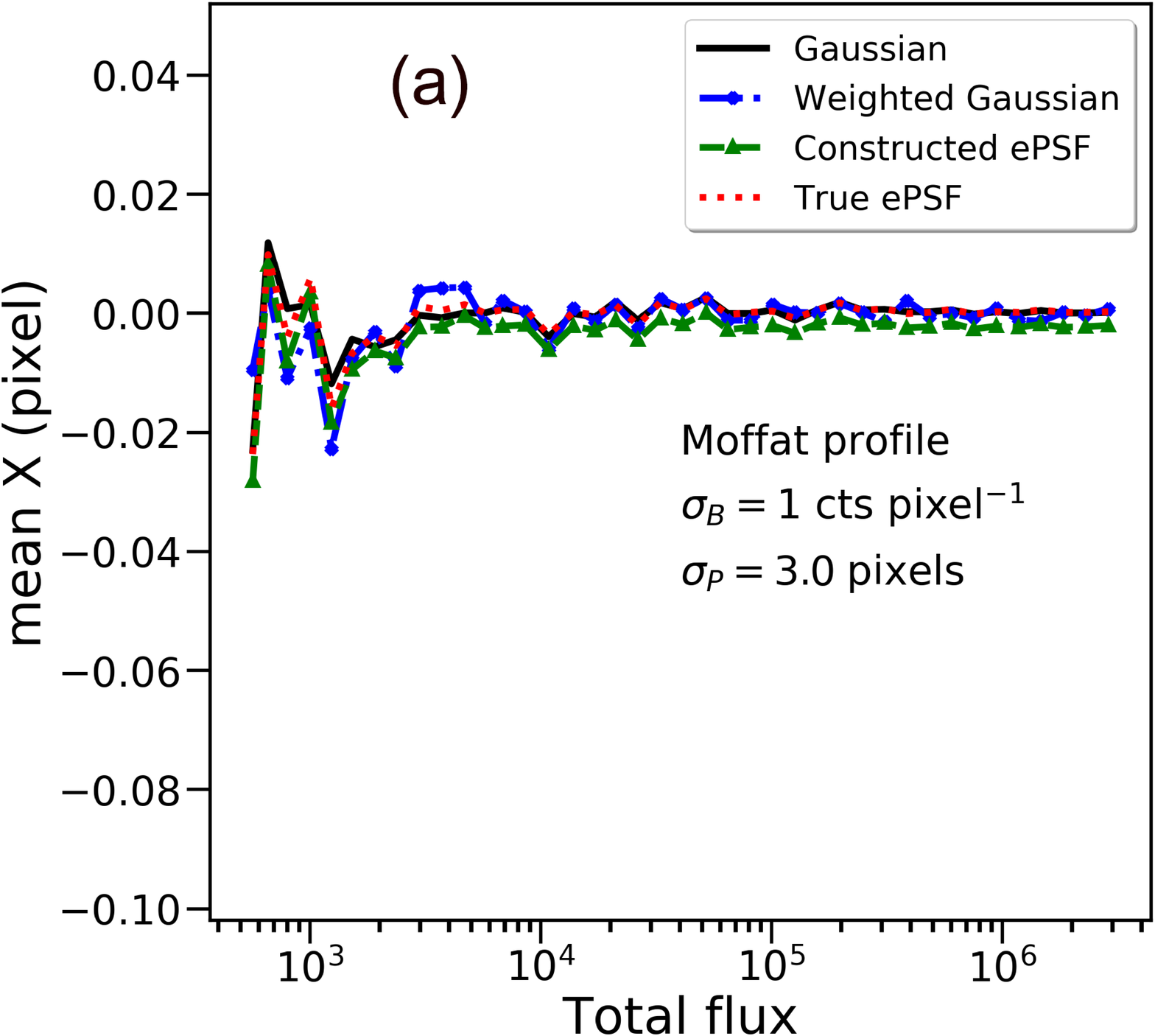}\
\includegraphics[width=0.42\textwidth]{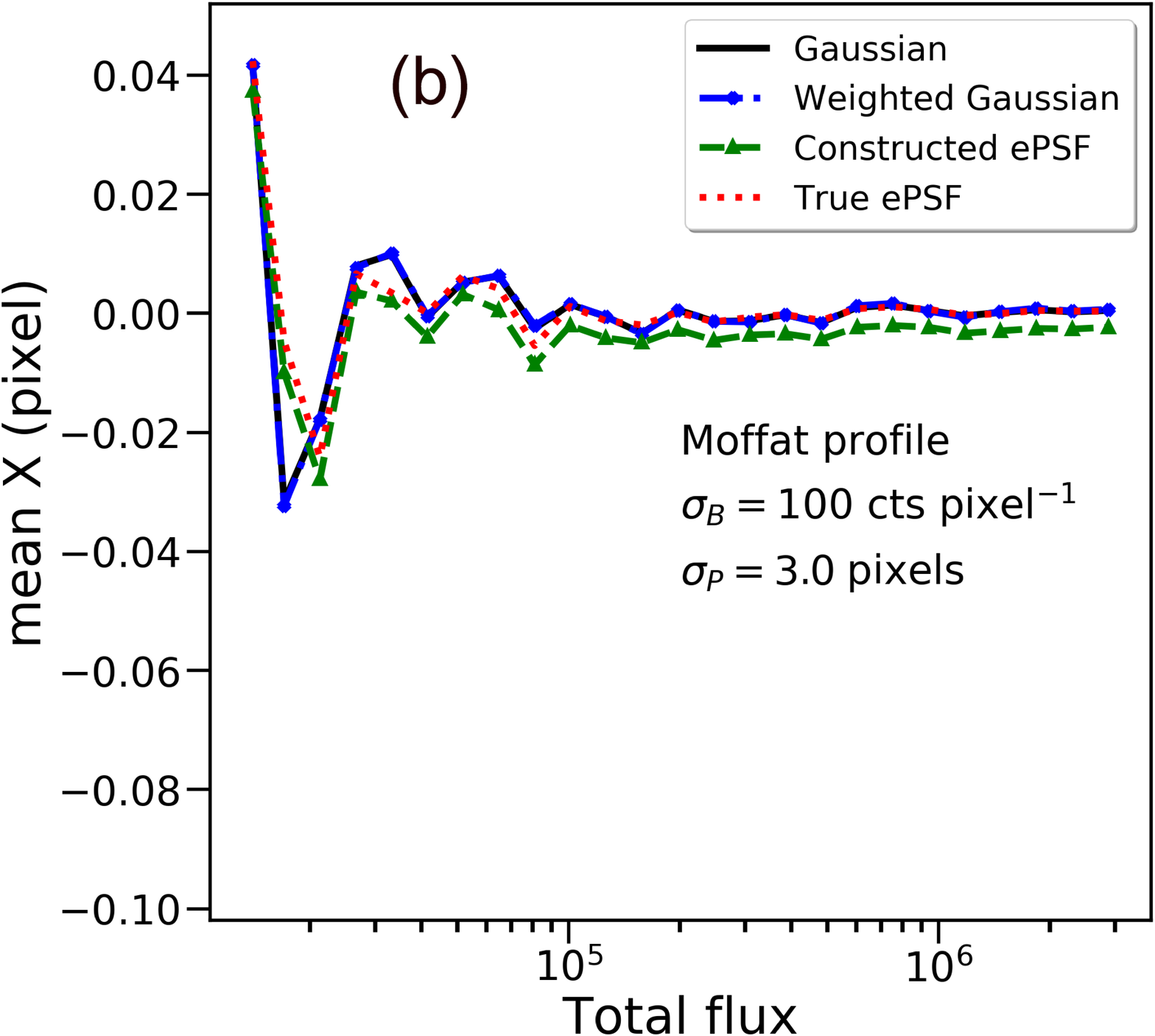}
\includegraphics[width=0.42\textwidth]{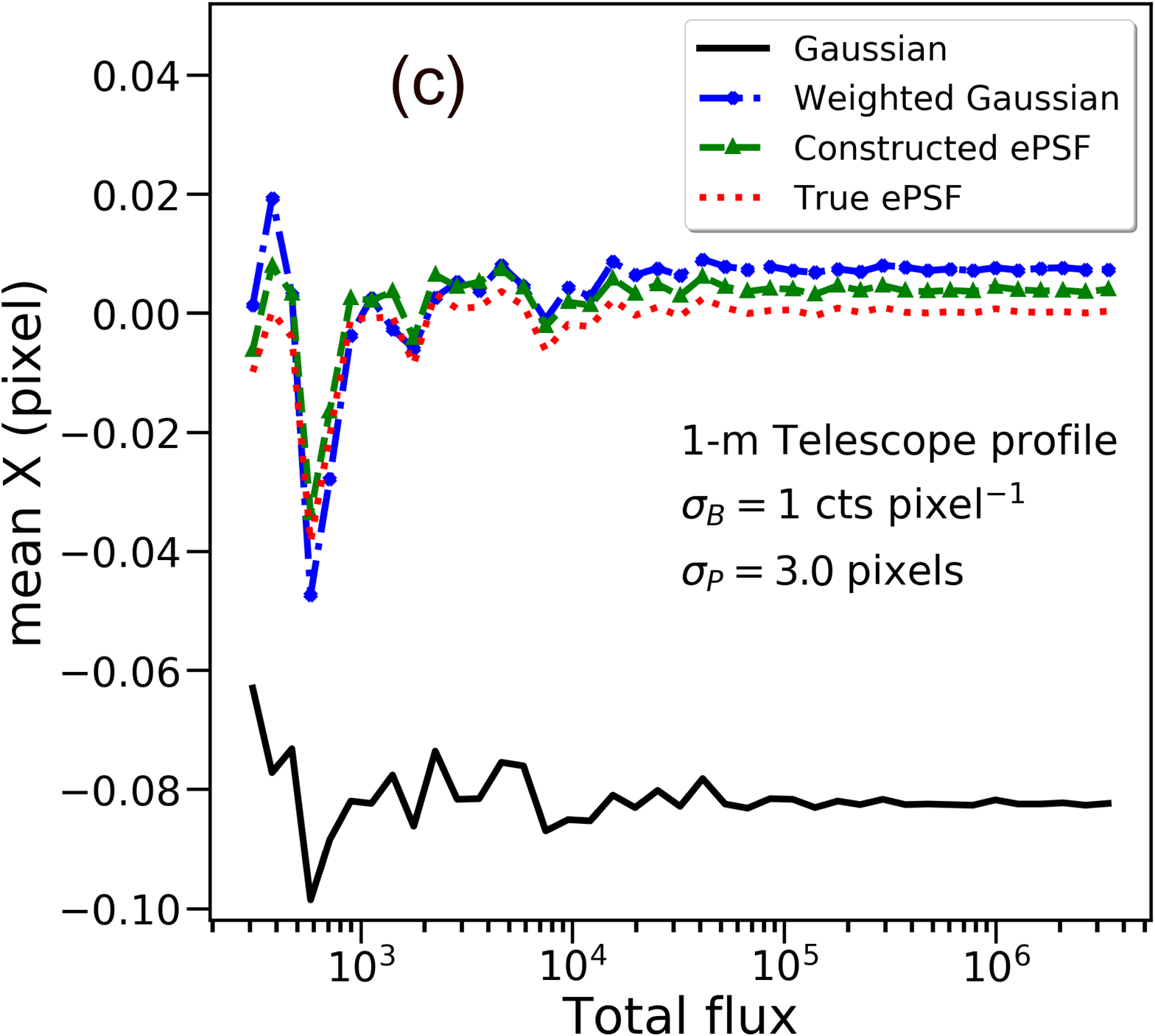}\
\includegraphics[width=0.42\textwidth]{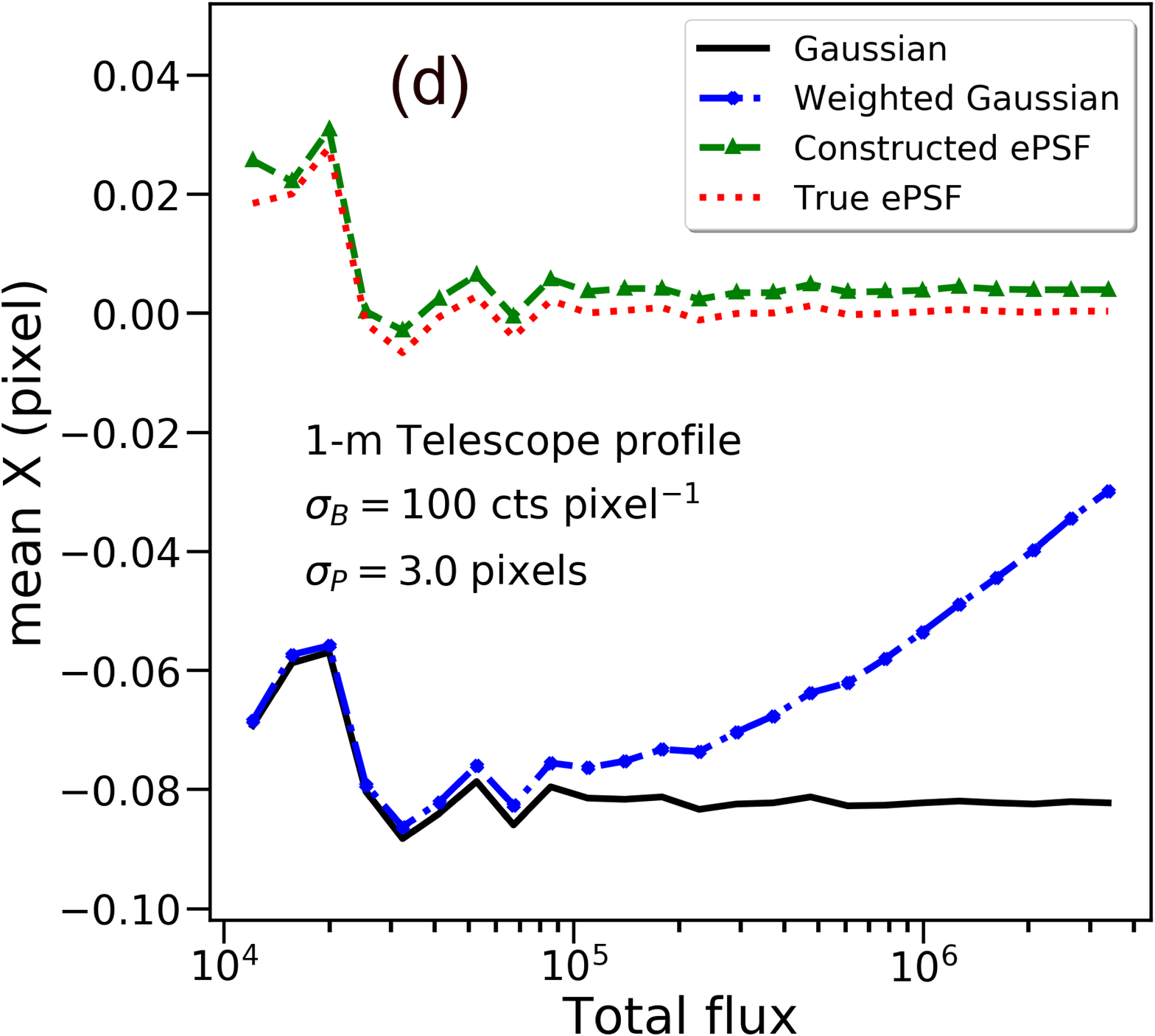}
		\caption{The mean residuals between the real and measured positions changing with the brightness of stars.} 	
	\label{fig7}
\end{figure*}

 Profiles of star images can affect the precision difference between the Gaussian and the ePSF fitting significantly. When the profile of star images has a sharp center, the precision of the ePSF fitting is almost the same as that of the Gaussian. The shape of practical profiles are between the extreme Moffat profile (Profile 2) and the Gaussian profile (as shown in Figure \ref{fig3}). Therefore, the precision improvement of the ePSF fitting for the simulation sets 3 and 4, whose profiles are extracted from observations, are in between (see Figure \ref{fig6}).

The weight of the least square fitting has a crucial effect on the centering precision. The Gaussian fitting simply uses unit weight , so it is theoretically not accurate enough. The weighted Gaussian fitting may have the same high precision as the ePSF fitting since they use the same weight. However, as shown in Figure \ref{fig6}, this is true only when the profile of the star image is the symmetric Gaussian, and the centering error of the weighted Gaussian fitting would be larger in other situations. This occurs for two reasons.
Firstly, the weighted Gaussian fitting introduces the systematic error associated with the brightness of stars when the profile is not symmetric and the background is not low. Figure \ref{fig7} shows the mean residuals between the real and measured positions changing with the brightness of stars. It can be seen from Figure \ref{fig7}(c) that for Profile 3, the mean residuals of the weighted Gaussian fitting is not equal to that of the Gaussian when the background is negligible. Furthermore, the weighted Gaussian fitting degenerates into the Gaussian fitting when the background dominates. Consequently, its results will gradually shift as the star brightness increases. The right panel of Figure \ref{fig8} shows the low precision when the systematic error exists. Note that the practical PSFs of ground-based observations are rarely absolutely symmetric, so this error would usually exist for the weighted Gaussian fitting. Nevertheless, this systematic error can be relieved by using a plate model with a magnitude term in the reduction process.
The other reason is that the weight used in the weighted Gaussian fitting is inaccurate when the profile of the star images deviates from Gaussian.
 \citet{espinosa2018optimality} makes a thorough analysis of the correct choice of the weights, and it is pointed out that the optimal weights are given by $1/\lambda_i$ ($\lambda_i$ is the true flux in pixel i). We know the optimal weights are unfeasible to get in advance, since we need the true-position of the star (which is a priori unknown!), however, they give us a good tool for analysis. For example, under the Gaussian assumptions, it would be expected that the observation is not very distinct from the optimal weight, this is, $I_i \sim \lambda_i$ ($I_i$ is the observed flux), so then we should expect a good performance. However, if the profile of the star images deviates from the Gaussian assumption, it makes sense that $I_i$ is not very close to $\lambda_i$, and then the performance is not necessarily optimal (i.e., not close to CR bound).
The lower performance is caused by an error similar to the pixel-phase error. This error is especially serious when the star profile has gentle wings (left panel of Figure \ref{fig3}). The low precision of the weighted Gaussian fitting in this situation can be seen from Figure \ref{fig6}(b).
That is to say, the weight should only be used when the fitting model (i.e. function $p$ mentioned in Section \ref{sec2}) is accurate.

As the dominance of background noise increases, the effect of the weight on centering precision is gradually reduced. In extreme case, Equations (\ref{eq6p}) will degenerate into Equation (\ref{eq8}). At this time, only the accuracy of the fitting model affects the centering precision. It can be seen from Figure \ref{fig8} that ePSF fitting does not improve the centering precision in this situation. That is to say, an accurate fitting model has little value without the advantage of the fitting weight.
\begin{figure*}	
	\centering
\includegraphics[width=0.42\textwidth]{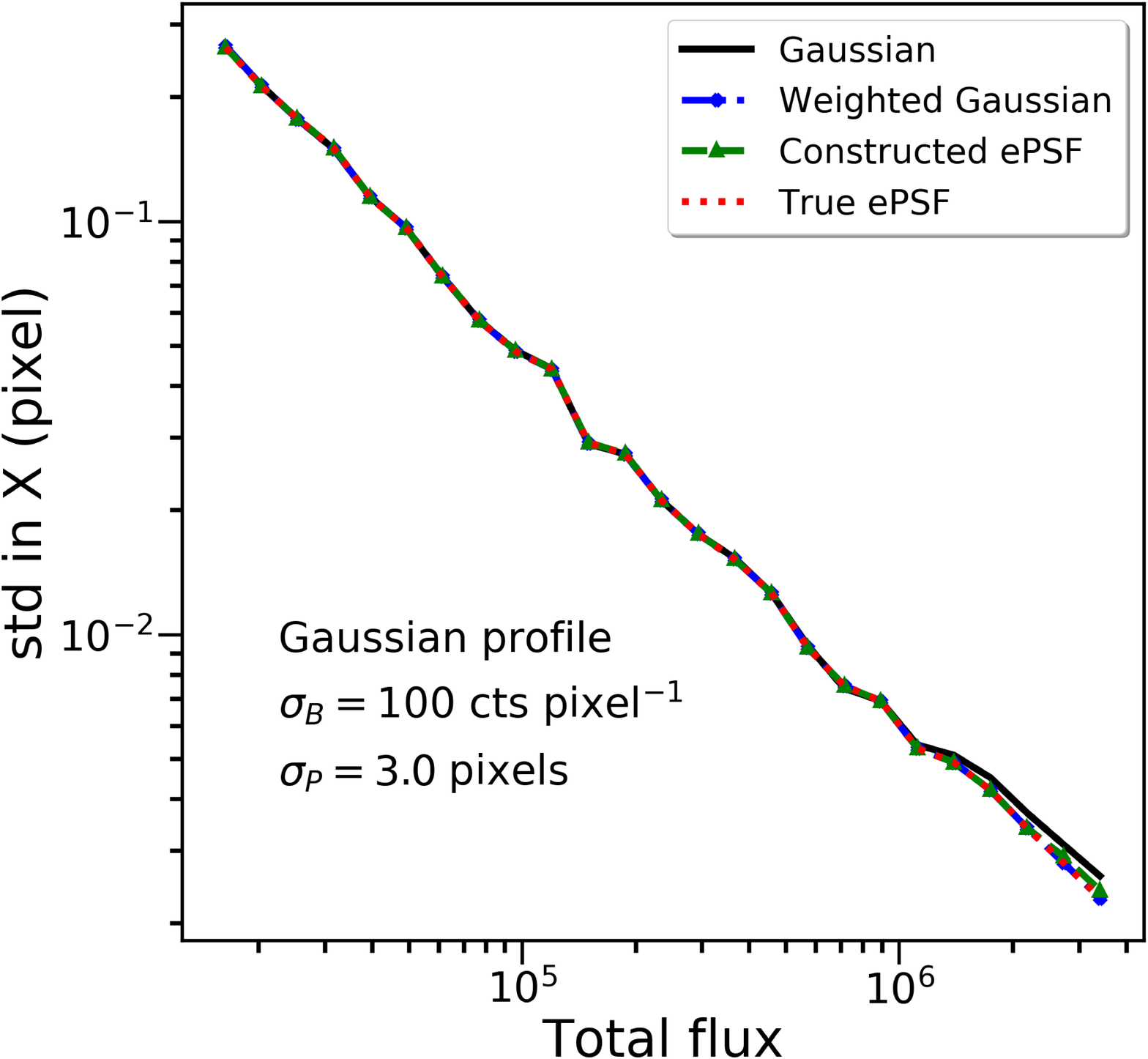}\
\includegraphics[width=0.42\textwidth]{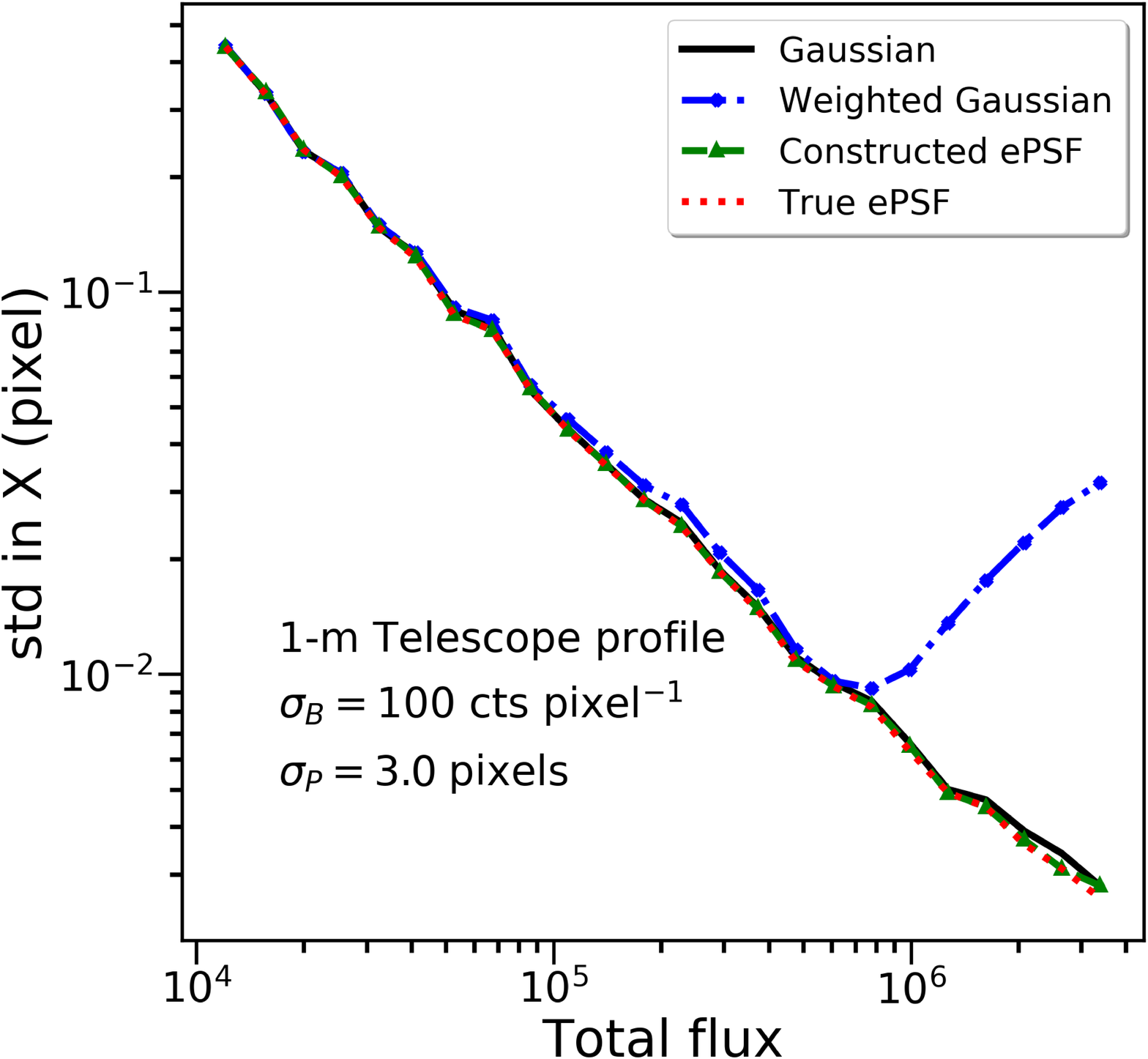}
		\caption{Centering errors of the synthetic images with different profiles when the background is extremely high.} 	
	\label{fig8}
\end{figure*}

The centering precision when the PSF changes with location on the detector is studied using the synthetic frames with Profile 5. Figure \ref{fig11} shows the standard deviation (std) of the original centering residuals, and the std of residuals after mapping the original results onto the real positions utilizing a plate model with a fourth-order polynomial. Before transformation, the centering precision of the Gaussian fitting is worse than that of the true ePSF fitting. The reason is that spatial variation of the PSF will lead to a systematic error related to the location on the detector. The lower precision of the constructed ePSF fitting is also caused by this systematic error, since there are re-centering blemishes during the process of the ePSF construction.
However, this error can be absorbed by a higher-order plate model when the PSF does not changes rapidly with location. Using the geometric distortion correction technique \citep[e.g.][]{peng2012convenient,wang2019distortion}, the case of the PSF changes rapidly with space can also be handled.
That is to say, the spatial variation of PSF itself does not introduce systematic error into the reduction results. The advantage of the ePSF algorithm at this time is that it can interpolate an accurate fitting model at each CCD position to fit the star image there.
Therefore, the final results when the PSF changes with location on the detector are similar to that of the constant PSF discussed above (see the right panel of Figure~\ref{fig11}).

\begin{figure*}	
	\centering
\includegraphics[width=0.42\textwidth]{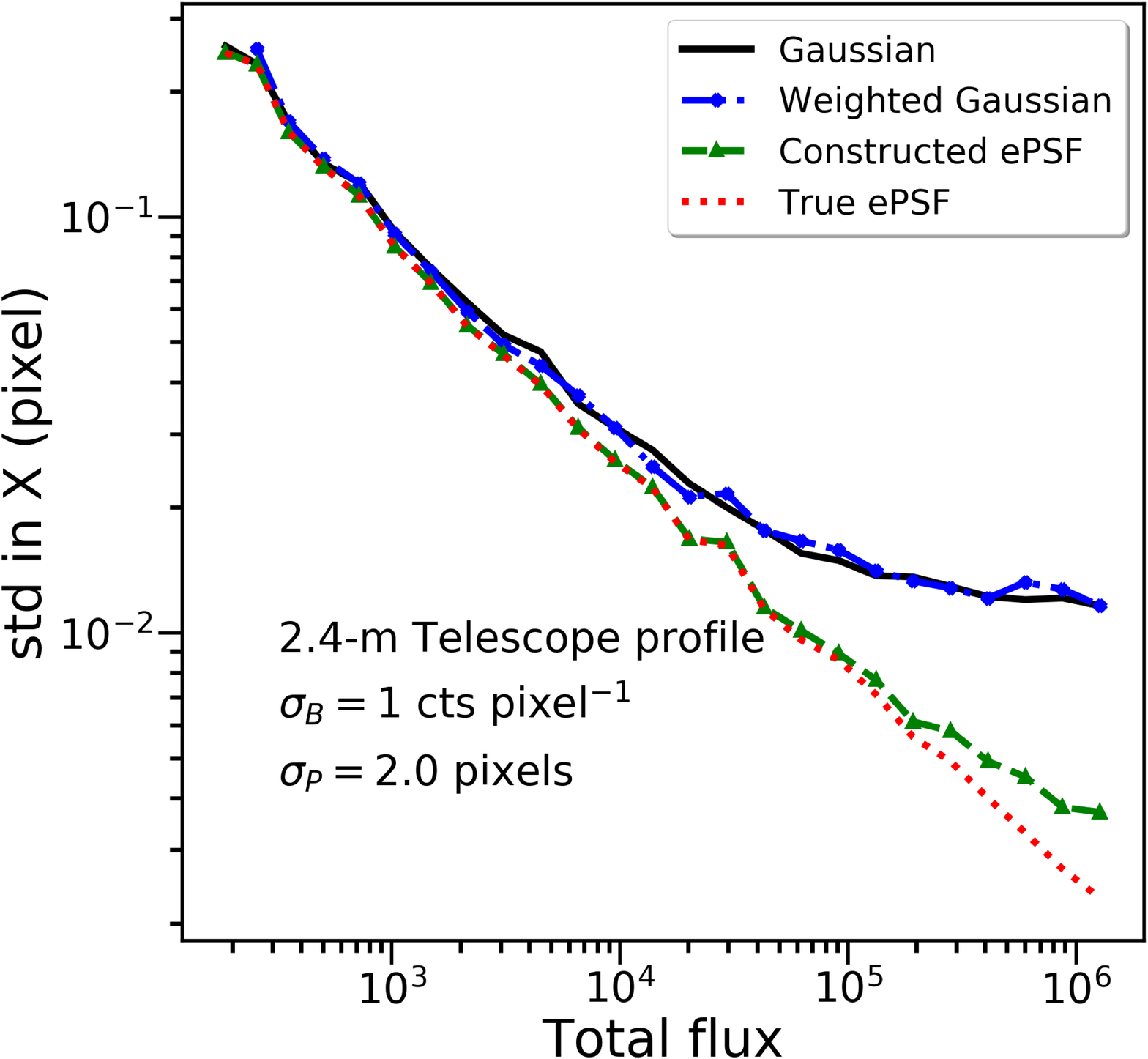}
\includegraphics[width=0.42\textwidth]{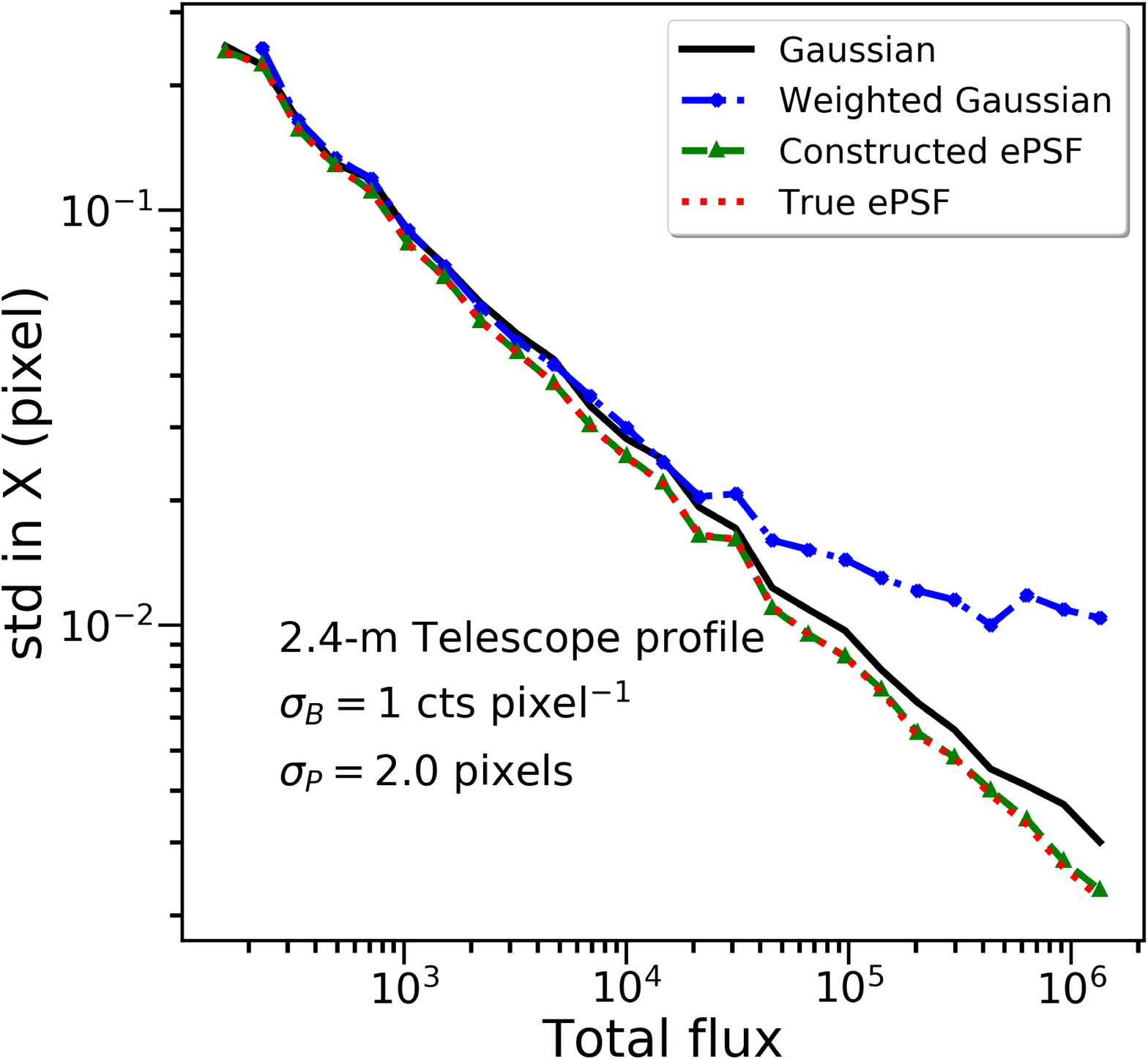}
		\caption{The std of the original centering residuals, and the std of residuals after a thirty-constant plate model transformation. } 	
	\label{fig11}
\end{figure*}

The pixel-phase error always exists for the undersampled images when the fitting model and the PSF of the star image unmatched.
 Figure \ref{fig10} shows the centering results of critical sampling star images (FWHM=2 pixels) and images with larger FWHM. It can be seen that with Gaussian fitting, it is difficult to obtain high-precision positions even for bright stars. The reason for this is that it leads to a large pixel-phase error for the critical sampling star images with the Moffat profile. The ePSF algorithm can construct an accurate ePSF from a single frame to avoid this error and obtain better results. To deal with the critical sampling star images, the ePSF algorithm only requires more sampling stars to satisfy finer pixel divisions (six grid points per pixel width in our simulation), and appropriate smooth to the ePSF model. For the profile used here, the pixel-phase error of Gaussian fitting does not disappear until FWHM $\ge$ 2.5 pixels.

\begin{figure*}	
	\centering
\includegraphics[width=0.42\textwidth]{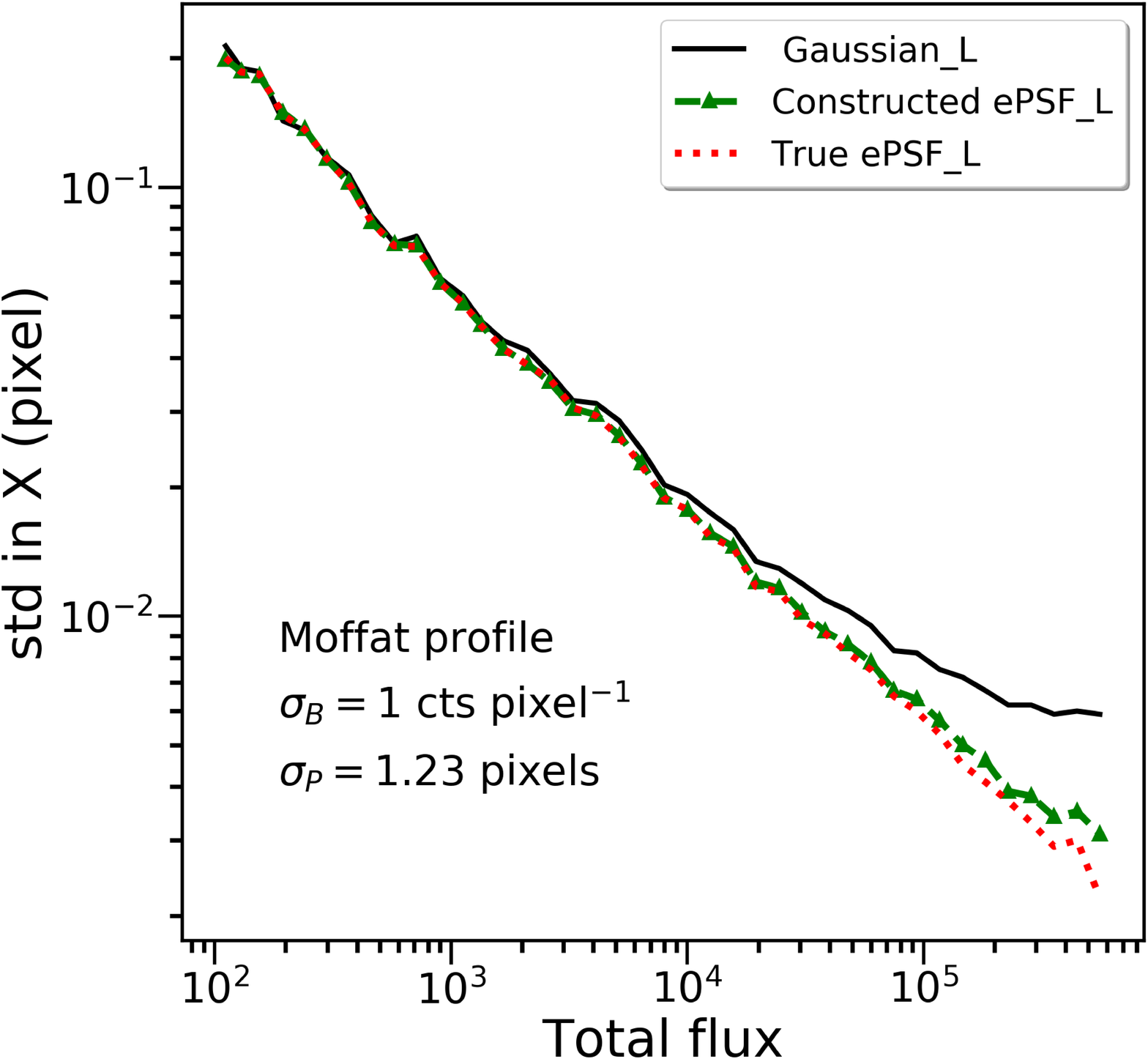}
\includegraphics[width=0.42\textwidth]{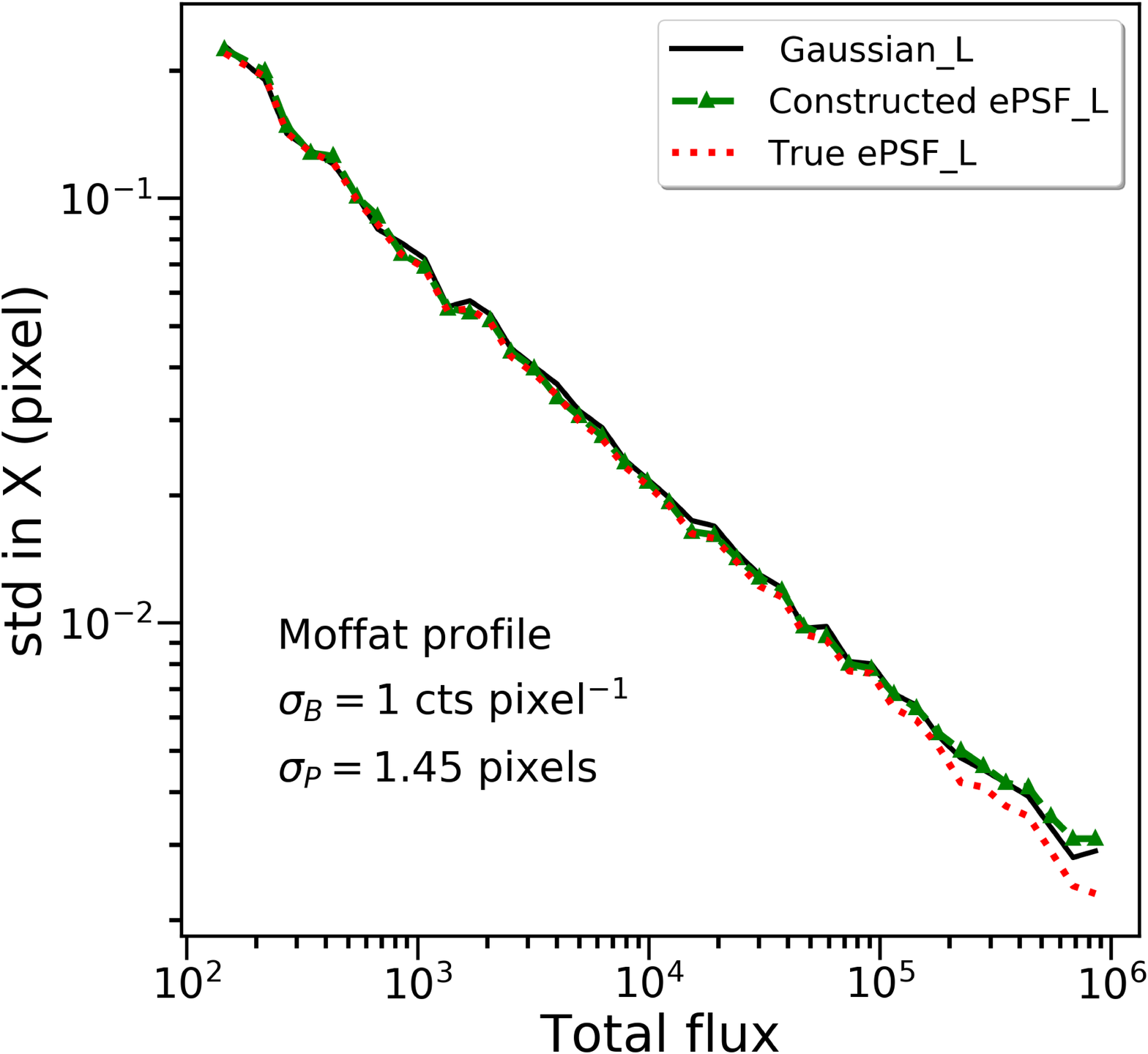}
		\caption{The simulation results for critical sampling star images (Moffat profile with $\beta=1.2$ and $FWHM=2$ pixels) and images with larger FWHM. For the profile used here, the pixel-phase error of Gaussian fitting does not disappear until FWHM $\ge$ 2.5 pixels.} 	
	\label{fig10}
\end{figure*}

\subsection{Results of the observations}
\begin{figure*}	
	\centering
\includegraphics[width=0.48\textwidth]{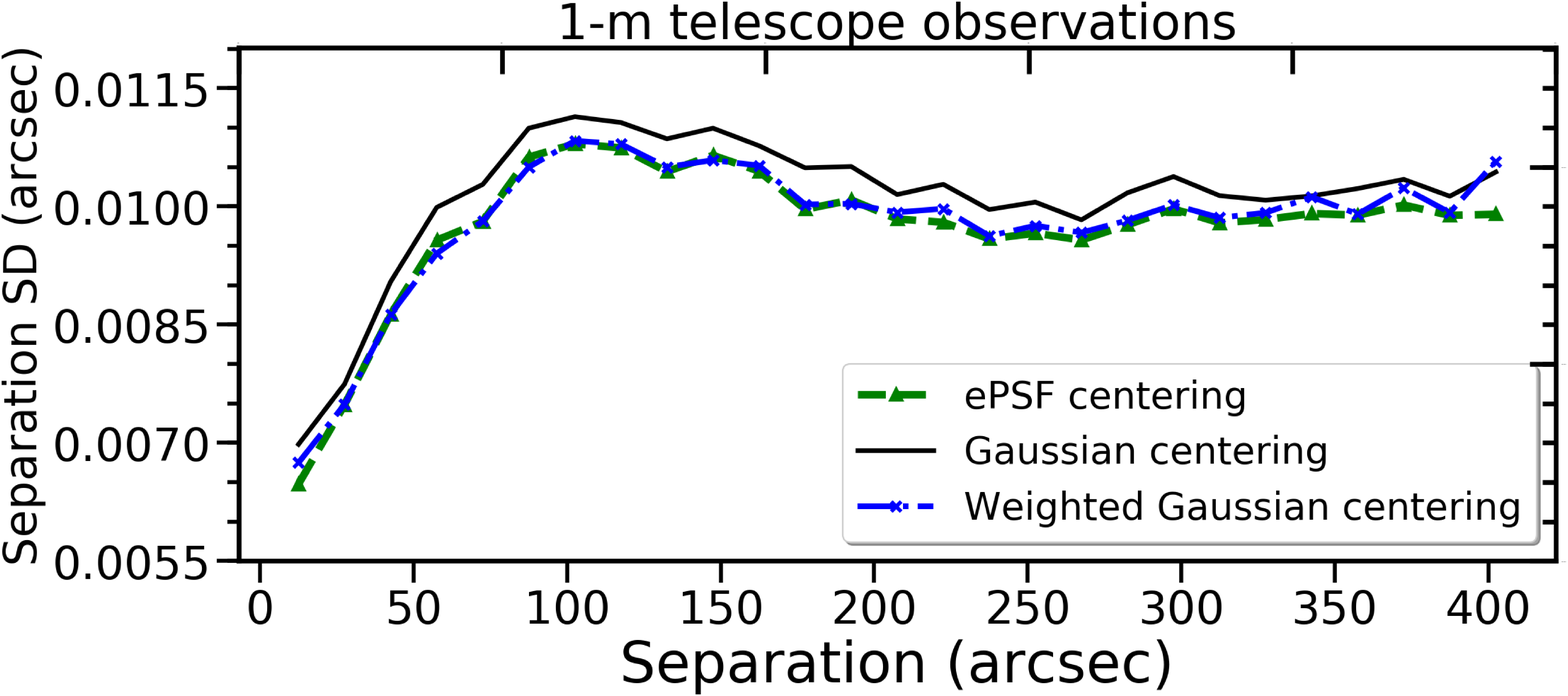}
\includegraphics[width=0.48\textwidth]{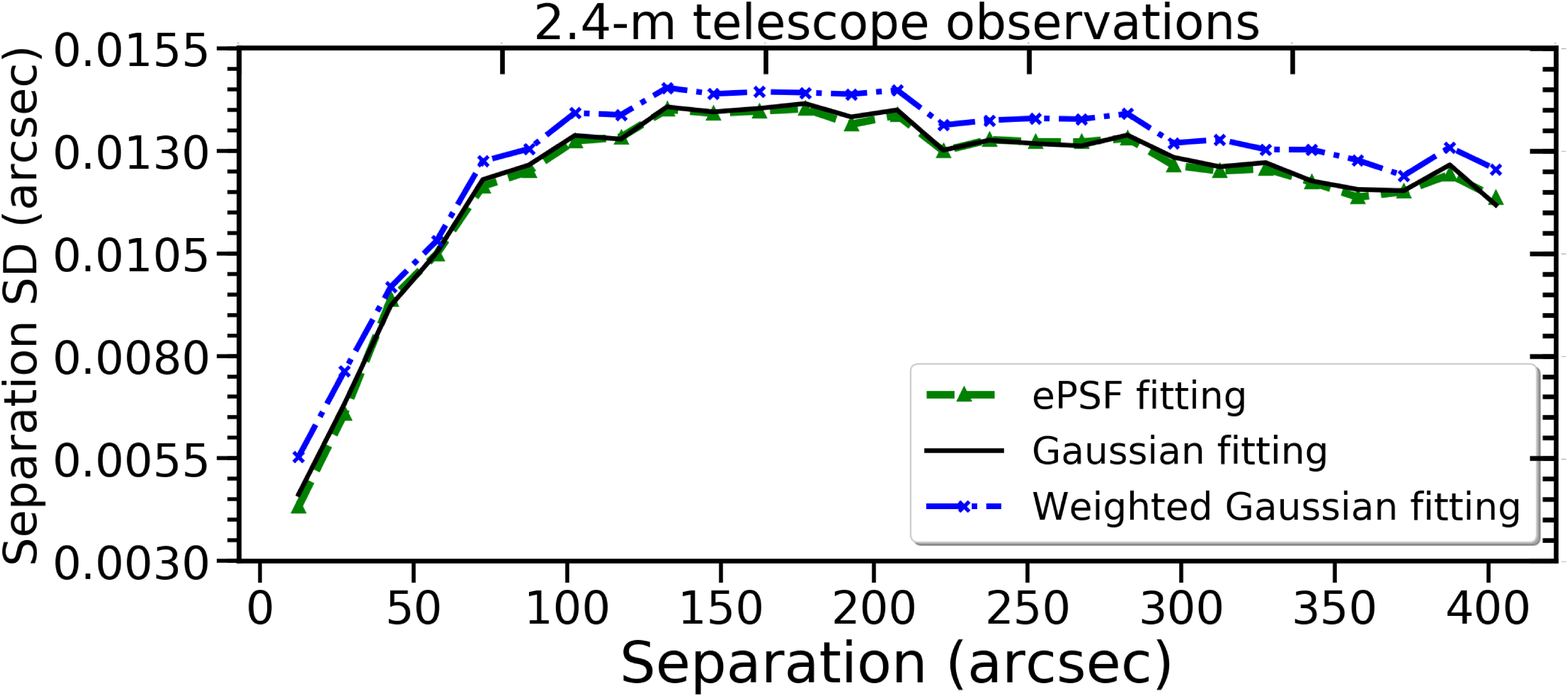}
		\caption{Comparison of the precision premium statistics for the observations processed by different fitting methods. The horizontal axis is the separation of the star pairs and the vertical axis is std of the separation.} 	
	\label{fig12}
\end{figure*}

Practical observations usually suffer from many kinds of errors, such as geometric distortion, the impact of atmospheric turbulence and differential color refraction. Some of them cannot be completely corrected in the data reduction process. Therefore, the small improvement of the centering precision may be submerged in other errors, and it is difficult to compare the precision of different centering algorithms by using observations. Fortunately, most of these errors can be minimized in the relative position measurement when the separation of two objects is small. The precision premium curve, which is generated by the statistics of the separation residuals, can clearly show the situation.
A detailed description of this curve is given in \citet{lin2019characterization}. After data reduction, the statistics of the separation residuals are given in Figure \ref{fig12}. Only stars brighter than 14 Gaia G-mag are used in the statistics.

It can be seen from Figure \ref{fig12} that the precision of the results with the method of the ePSF fitting is higher than that of the Gaussian. For our 1-m telescope observations, the separation std decreases from about 7.0 mas to 6.5 mas. Namely, the centering precision is improved by about 2 mas when the ePSF fitting used. The improvement is merely about 1 mas for our 2.4-m telescope observations. The less improvement of precision for the 2.4-m telescope results is probably due to the high background of the observations, which reduces the effect of the weight in the ePSF fitting.

For 1-m telescope observations, the results of the weighted Gaussian fitting are comparable to that of the ePSF fitting. However, the conditions for using it to obtain benefits are very demanding. Just as the 1-m telescope observations, it needs a low background, and the ePSFs should be close to the Gaussian (see parameter $\beta$ in Table~\ref{tbl2}). Moreover, most of the star images in each frame should have a medium or low SNR, only a few have a high SNR.
For the 2.4-m telescope observations, the weighted Gaussian fitting has obvious lower precision than other methods.
Therefore, weighted Gaussian fitting should be used with caution in the processing of practical observations.
\section{Summary and conclusions}
In this paper, the precision difference between centering algorithms based on model fitting was studied through a detailed comparison of the ePSF and the Gaussian centering algorithms.

Using an accurate fitting model and appropriate weight to fit star images, we can obtain the highest centering precision. The ePSF algorithm can construct the accurate ePSF to attain this high precision.
The Gaussian centering algorithm could also achieve the same precision under suitable observational conditions, but it would show slightly lower precision when it is not used properly.
Specifically, when the PSF is close to the Gaussian, the precision of the ePSF fitting can be higher than that of the Gaussian fitting from medium to high SNR. As the theoretical prediction given in \citet{lobos2015performance} and the simulations shown in Figure~\ref{fig6p}, the improvement can be up to $24.1\%$ under ideal conditions. However, when the center of the ePSF becomes sharper, the precision of the Gaussian fitting and the ePSF fitting tends to be the same. This is also the case when the SNR of star images is decreased.
The weighted Gaussian fitting, which uses the same weight as ePSF fitting, could have the same advantages as the ePSF fitting when the observations have the Gaussian PSF, medium SNR and the low background (as our observation set 1). But it turned out to be poor when the PSF has a certain difference with the Gaussian.

The centering precision when the shape of the PSF changes with location on the detector was also investigated. The spatial variation of the PSF results in a systematic error related to the location, but it could be absorbed in the reduction. As a result, the advantages of the ePSF algorithm in this situation are the same as those in the situation of constant PSF.
The PSF usually change with CCD frames for the ground-based observations \citep{stetson1987daophot}, so the effect (amplitude or phase) of pixel-phase error would also change with frames. Our simulations showed that an accurate ePSF can be constructed from a single CCD frame in which the star images have small FWHM ($\sim$2 pixels). Therefore, the ePSF algorithm can handle the pixel-phase error even for the ground-based observations.
For the observations with few stars in the field of view, the model error of the constructed ePSF will be large and lead to the degradation of the ePSF algorithm. In this paper, we demonstrated that the best performance can be achieved when the number of sampling stars is more than $\sim$15 times of grid points in a pixel.

Since the performance of each algorithm in practical application was clarified in detail, the appropriate algorithm can be selected according to the image characteristics of observations. The ePSF algorithm shows higher performance when there are enough sampling stars to construct the accurate ePSF. But, the Gaussian centering algorithm is easy to use (less resources are needed), and is generally better, with a degradation in performance that is not necessarily too harmful (10\% in our practical application). Moreover, the weighted Gaussian fitting could have as good results as the ePSF algorithm when the assumptions are more consistent with the Gaussian scenario, and its performance would degrade only if the profile deviates from the Gaussian assumption.

In order to verify the conclusions, two sets of observations obtained from the 1-m and 2.4-m telescopes at Yunnan Observatory were reduced using the algorithms above. When the algorithm with the highest precision was adopted, the precision for these two observation sets was improved by about 2 mas and 1 mas respectively compared with the Gaussian centering algorithm.

 \acknowledgments
This work was supported by the National Natural Science Foundation of China (Grant Nos. 11873026, 11273014), by the Joint Research Fund in Astronomy (Grant No. U1431227) under cooperative agreement between the National Natural Science Foundation of China (NSFC) and Chinese Academy Sciences (CAS), and partly by the Fundamental Research Funds for the Central Universities. We would like to thank the chief scientist Qian S. B. of the 1-m telescope and his working group for their kindly support and help. We also thank the referee for the insightful review of our manuscript, and thank Dr. Nick Cooper from Queen Mary University of London for his linguistic assistance during the preparation of this manuscript. This work has made use of data from the European Space Agency (ESA) mission \emph{Gaia} (\url{https://www.cosmos.esa.int/gaia}), processed by the \emph{Gaia} Data Processing and Analysis Consortium (DPAC, \url{https://www.cosmos.esa.int/web/gaia/dpac/consortium}). Funding for the DPAC has been provided by national institutions, in particular the institutions participating in the \emph{Gaia} Multilateral Agreement.


%
\bibliographystyle{spr-mp-nameyear-cnd}  
\bibliography{bibtex}

\end{document}